\documentstyle[aps,manuscript,epsf,eqsecnum]{revtex}  

\newcommand{\eqn}[2]{\begin{equation}\label{#1} #2 \end{equation} }
\newcommand{\avg}[1]{\left\langle #1 \right\rangle }
\newcommand{\D}[2]{ \delta^d\left(\vec{\xi}(#2) - #1\right) }

\begin{document}
\title{ Anomalous Fluctuations of  \\
Directed Polymers in Random Media }
\author{Terence Hwa~\cite{addr} and Daniel S. Fisher}
\address{Lyman Laboratory of Physics, Harvard University, Cambridge, MA 02138}

\date{\today}
\maketitle
\widetext
\begin{abstract}

A systematic analysis of large scale fluctuations in the low temperature 
pinned phase of a directed polymer in a random potential is described.
These fluctuations come from rare regions with nearly degenerate 
``ground states''. The probability distribution of their sizes is found 
to have a power law tail. The rare regions in the tail dominate much of the
physics. The analysis presented here takes advantage of the mapping to 
the noisy-Burgers' equation. It complements a phenomenological
description of glassy phases based on a scaling picture of droplet excitations
and a recent variational  approach with  ``broken replica symmetry''. 
It is argued that the power law distribution of large thermally active 
excitations is a consequence of the continuous statistical ``tilt'' symmetry 
of the directed polymer, the breaking of which gives rise to the large 
active excitations in  a manner analogous to the appearance of 
Goldstone modes in pure systems with a broken continuous symmetry.

\end{abstract}


\section{INTRODUCTION}

The statistical mechanics of directed polymers in random media has
attracted much attention in recent years~\cite{hh,hhf,iv,kz,mm}. 
This problem and related problems of higher dimensional manifolds are 
encountered in
a variety of contexts, ranging from the fluctuations of domain
walls in random
magnets~\cite{hh,dw,manifold}, to the dynamics of magnetic
 flux lines in dirty superconductors~\cite{sc,vg}. In addition,
the randomly-pinned directed polymer is one of the simplest models that
contain many of the essential features of strongly frustrated
random systems such as spin-glasses~\cite{sg}. Understanding the 
 behavior of the directed polymer is therefore important
for developing intuition and testing
theoretical ideas for more complicated random systems.

Over the years, a variety of methods have been used
 to study random directed polymers. These include mapping~\cite{hhf,iv}
 to a hydrodynamic system: the 
noise-driven Burgers equation~\cite{FNS,KPZ}, 
 the exact solution on a Cayley tree~\cite{cayley},
Migdal-Kadanoff approximate renormalization group calculations~\cite{migdal},
a Bethe ansatz solution in 1+1 dimensions using replicas~\cite{bethe,p},
a gaussian variational ansatz in  replica space~\cite{mp},  
and finally, renormalization group arguments and phenomenology~\cite{fh} 
in the spirit of  the droplet (or scaling) theory of 
spin-glasses~\cite{droplet}.
There have also been substantial numerical simulations;
some recent studies can be found in Refs.~\cite{fh,m,kbm,kmh}.
The qualitative phase diagram of the directed polymer
 is found to be quite simple:
  There is always a pinned phase dominated by disorder at 
low temperatures.
For polymers in $d+1$ dimensions, only the pinned phase 
exists for $d\le 2$. But
for $d > 2$, the polymer can undergo a continuous transition
to a high-temperature phase where the disorder 
is irrelevant as has been proved
rigorously~\cite{is}.
Until recently, most of the efforts have been focused on
characterizing the scaling properties of the polymer displacements and
free energy fluctuations in the pinned
phase. With the exception of 1+1 dimensions for which the scaling exponents
can be computed exactly, systematic and analytic computations of the 
exponents at the zero temperature fixed point that controls the pinned phase
have not been possible so far. 

Some of the other properties of the glassy pinned phase
beyond the scaling exponents were explored numerically
by Zhang~\cite{z}, and more recently by M\'{e}zard~\cite{m}. 
These authors find
very sensitive dependence of the polymer's low temperature
configuration on the details of the particular random medium.
Such sensitivity is associated with rare but singular dependence on
 the details of the random potential of the system.
This type of behavior, including sensitivity to small temperature changes,
 has  been predicted by Fisher and Huse~\cite{fh} by phenomenological 
scaling and renormalization group arguments.
As argued in Ref.~\cite{fh} (and supported by numerical simulations in
\cite{fh,m,z}), the physics of the low temperature phase is dominated by large
scale, low energy excitations of rare regions, analogous to the droplet
excitations proposed for Ising spin glasses~\cite{droplet}. 
 Although the picture developed there
is physically appealing and relatively complete, the phenomenological 
approach of Ref.~\cite{fh}
does not provide a systematic or quantitative
 way of calculating the properties of  the pinned phase. 
On the other hand, various uncontrolled approximations 
can provide quantitative information. In
particular, the Migdal-Kadanoff calculations of Derrida and
 Griffiths~\cite{migdal} could be used to study the properties predicted 
in Ref.~\cite{fh} although this
has not been carried out in detail. M\'{e}zard and Parisi~\cite{mp} 
have recently 
 proposed a very different approach: a variational method in replica space
 which can be used to study various aspects
of the pinned phase.  However, the method is limited by the gaussian Ansatz
used whose physical significance is not clear, 
and the analysis based on replicas is haunted by the usual problem of the 
interchange of the $n\to 0$ and the thermodynamic limits.
This is particularly problematic because within the  gaussian Ansatz, 
one finds that the solution selected 
 requires broken replica symmetry, the physical interpretation 
of which (if any) is unclear. (Note that the {\it correct} scaling 
exponents were obtained
in 1+1 dimensions by using the  Bethe Ansatz in replica space 
{\it without} breaking replica symmetry~\cite{bethe}.)

In this paper, we explore the properties of the pinned phase 
of random directed
polymers by using a more conventional approach based on a field-theoretic
description without replicas, and the statistical symmetries of the problem.
We will see that the {\it existence} of scaling forms for 
long wavelength, low frequency
correlators of the noisy-Burgers' equation {\it implies}, 
without significant additional
assumptions, the existence of rare large scale, low energy 
``droplet'' excitations
with a power-law distribution of their sizes. Although the 
large thermally active
excitations are very rare, they dominate many thermodynamic 
properties and average
correlation and response functions, as well as causing 
large variations in the properties
of macroscopic systems.   We therefore see that many of the properties
of the pinned phase predicted by Fisher and Huse~\cite{fh} 
and found in other uncontrolled approximations can be 
recovered from the existence of a fixed point
which, in the hydrodynamic language, is rather conventional. 
A fundamental lesson from
this is that the broken continuous statistical symmetry of 
the pinned phase of the
directed polymer gives rise, quite generally, to power law 
distribution of large rare,
low energy excitations; these are  the analog for broken continuous
statistical symmetries of the Goldstone modes associated 
with true broken  continuous symmetries !

This paper is presented in a somewhat pedagogical fashion. 
It is intended to introduce the ideas of rare fluctuations as well as 
providing an alternative perspective for those familiar 
with the scaling approach
to directed polymers and spin glasses. The paper is organized as follows:  
In Section II, we define the directed polymer problem
and some of the glassy properties of the pinned phase. 
We motivate the considerations of almost degenerate ground states 
that give rise to large scale,
 low energy excitations, and relate their statistics
to the distribution functions of the end point of a polymer.
The distribution functions are computed in
Section III by using a free-energy functional 
(described in Appendices A and B)
and by exploiting the statistical symmetries (Appendix C). 
The results are interpreted in terms of the large,
rare fluctuations in the pinned phase, with a
short discussion contained in Section~IV. 
The results of this study are then compared with
the variational approach of Ref.~\cite{mp} 
and the phenomenological approach
of Ref.~\cite{fh}. We conclude that the various approaches point to the same
physics governing the excitations of the directed polymer
  at low-temperatures --- the physics of large, rare fluctuations.

\newpage
\section{ PROPERTIES OF THE PINNED PHASE}

\subsection{The Model}

We consider a directed polymer of length $t$ (which will later be convenient
to consider as ``time'') embedded in $d+1$
dimensions. Let the position vector $(\vec{\xi}(z),z)$ with 
$\vec{\xi} \in \Re^d$ and $0\le z\le t$ describe the
path of the polymer
 in the $d$ transverse dimensions. Then the statistical
mechanics of this polymer is determined by the Hamiltonian
\eqn{H}{ {\cal H}[\vec{\xi},\eta] =  \int_0^t dz \left[ \frac{\kappa}{2} 
\left(\frac{d\vec{\xi}}{dz}\right)^2 + \eta(\vec{\xi}(z),z)\right], }
where $\kappa$ is the line tension, and $\eta$
is a quenched random potential of the medium 
through which the polymer passes. The
random potential can be taken to be uncorrelated and gaussian
distributed, with $\overline{\eta}=0$ and
\begin{equation}
\overline{\eta(\vec{x},z)\eta(\vec{x}',z')} =
2D\delta^d(\vec{x}-\vec{x}')\delta(z-z'). \label{noise}
\end{equation} 
Here, the overbar denotes averages over $\eta$, and $D$ characterizes the
strength of the random potential. 
Note that a cutoff on the spatial delta function is
necessary for $d \ge 2$.

In order to make our subsequent discussions precise, it is useful to fix
one end of the polymer at an arbitrary point, say at $\vec{\xi}(t)=\vec{x}$. 
 This is implemented by introducing the one-point restricted
partition function,
\eqn{Z}{Z(\vec{x},t) = \int{\cal D}[\vec{\xi}] \> \D{\vec{x}}{t} 
e^{-{\cal H}[\vec{\xi},\eta]/T},}
where the arguments of $Z$ give the coordinate of the fixed end point. Unless
otherwise indicated,  thermal averages will be performed using this restricted
partition function, and will be denoted by $\avg{\dots}$.

The partition function $Z(\vec{x},t)$ is, of course, a random variable because
of its dependence on the realization of the random 
potential $\eta$.  For a particular
realization of $\eta$, the Hamiltonian has no symmetries. 
However, because of the
translational invariance in both $\vec{x}$ and $t$ of the 
{\it statistics} of $\eta$,
the {\it distribution} of ${\cal H}$ has {\it statistical 
translational symmetry} in $\vec{x}$
and $t$. Thus the statistics of $Z$ and other properties will also be
translationally invariant.  As we shall see, this symmetry and the closely 
related ``tilt'' symmetry
--- which corresponds to a weakly $t$-dependent translation 
in $\vec{x}$ --- have
dramatic consequences.

One of the simplest way to characterize the configuration 
of the polymer is the
one-point distribution function:
\eqn{P11}{P_1(\vec{y},t_0|\vec{x},t) \equiv
\avg{\D{\vec{y}}{t_0}}.}
This function is the conditional probability of finding the segment,
$\vec{\xi}(t_0)$,
 of the polymer at position $(\vec{y},t_0)$ given that the fixed end
 is at $(\vec{x},t)$. (We use ``$|$" to separate the
positions of the free$|$fixed positions of the polymer as 
in conditional probability.)
 In the absence of disorder, this distribution is
easily calculated, yielding the usual random-walk result
\begin{equation}
G^{(0)}(\vec{x}-\vec{y},t-t_0)\equiv P_1^{(0)}(\vec{y},t_0|\vec{x},t) = 
 \left(\frac{\kappa}{2\pi T (t-t_0)}\right)^{d/2} \exp\left[-\frac{\kappa}{2T}
\frac{(\vec{x}-\vec{y})^2}{t-t_0}\right],
\end{equation}
where the superscript $(0)$ denotes the pure system with $D=0$.
 The second moment of this
distribution gives the mean transverse displacement,
\begin{equation}
\avg{|\vec{X}(t-t_0)|^2}^{(0)} \equiv 
\avg{|\vec{\xi}(t)-\vec{\xi}(t_0)|^2}^{(0)} 
\equiv \int d^d\vec{r} \,\vec{r}^2\, G^{(0)}(\vec{r},t-t_0) 
= \frac{Td}{\kappa} (t-t_0).
\end{equation}

The existence of the random potential tends to 
increase the transverse wandering of
the polymer, as it tries to take advantage of favorable regions of 
the random potential at low temperatures. After averaging over the disorder,
translational symmetry in space is restored, and we have
$\overline{P}_1(\vec{y},t_0|\vec{x},t) = G_t(\vec{x}-\vec{y},t-t_0)$.
Note that a subscript $t$ is used to indicate the explicit dependence of $G$
 on the polymer length; we have not put this subscript on
 $P_1$ since it already has the
explicit $t$ dependence.  Numerically, it is found that 
\begin{equation}
\overline{\avg{|\vec{X}_t(\tau)|^2}}\equiv \int d^d\vec{r} \,\vec{r}^2\, 
G_t(\vec{r},\tau) = \tau^{2\zeta} f(\tau/t),\label{moment1}
\end{equation}
with $f(s)$ being a smooth scaling function which is finite throughout 
the range $0\le s \le 1$, and
 the wandering (or roughness) exponent is $1/2 < \zeta < 3/4$ depending on
the dimensionality $d$~\cite{bound}. In 1+1 dimensions, 
the exponent $\zeta= 2/3$ 
has been obtained exactly by a number of methods~\cite{hhf,bethe}. 
The power law  dependence of the transverse displacement
reflects the lack of intrinsic scales. This is possible because of the
statistical translational symmetry, $\vec{\xi} \to \vec{\xi} + {\rm const}$
of the Hamiltonian (\ref{H}) as discussed above. In dimensions $d>2$, there is
a high temperature or weak randomness phase 
in which $\zeta=1/2$, in addition to
the low temperature pinned phase on which we shall focus.
 It has been conjectured
\cite{fh} that $\zeta > 1/2$ for all finite $d$ in the pinned phase 
although this is controversial.

The detailed form of the disorder-averaged end point function $G_t$ is much
harder to obtain than the scaling exponent, 
even numerically~\cite{kbm,kmh}. It is expected to have the general form 
\eqn{P}{G_t(\vec{r},\tau) \approx \tau^{-d\zeta} \widetilde
g_{t/\tau}(r/\tau^\zeta)} 
by simple scaling and normalization requirements, with the
scaling function $\widetilde g$ depending only weakly on $t/\tau$ but 
decaying rapidly for $r\equiv |\vec{r}| \gg \tau^\zeta$. 
Two limits of particular interests are 
\begin{equation}
 G(\vec{r},\tau) \equiv \lim_{t/\tau \to \infty} G_t(\vec{r},\tau)
\approx \tau^{-d\zeta}\widetilde g_\infty(r/\tau^\zeta), \label{Pinfinite}
\end{equation}
which describes the one-point distribution of a semi-infinite polymer,
and
\begin{equation}
\widetilde G(\vec{r},t) \equiv  G_t(\vec{r},t)
\approx t^{-d\zeta}\widetilde g_1(r/t^\zeta), \label{Pfinite}
\end{equation}
which describes the distribution of position of 
the free end of a finite polymer.
As explained in the Appendices, analytical studies of the problem is much
simplified in the limit 
\linebreak $t/\tau \to \infty$.
 For instance, $G(\vec{r},\tau)$
 is given to a good approximation by
 a self-consistent integral equation in 1+1 dimensions
 (See Appendix B and Ref.~\cite{HF}).
 Numerical solution of the integral equation~\cite{HF} yields the form
Eq.~(\ref{Pinfinite}), with $\widetilde g_\infty(s)$ well approximated 
by a gaussian, although the precise shape of the tail is not known.
However, in numerical simulations, it is most 
convenient to study the end-point
distribution $\widetilde G(\vec{r},t)$. It is found~\cite{kbm,kmh} that the
distribution function obeys the scaling form Eq.~(\ref{Pfinite}). 
A comparison between $\widetilde g_\infty(s)$ and $\widetilde g_1(s)$ shows  
that the scaling function at the two limits are quite similar:
both are sharply decreasing functions whose widths are of order unity.
Thus we see that the semi-infinite polymer result
 $G(\vec{r},\tau)$  gives the 
correct {\it scaling} behavior and the qualitative form of the scaling 
function for the end-point distribution of a finite polymer.
We shall use this approximation in the following sections, where
we will compute various distribution functions 
for a semi-infinite polymer (i.e.,
for  $\tau \ll t$) and then apply them to the end point of a 
finite polymer with $\tau = t$.

\subsection{Ground States}
\nopagebreak
The description of the directed polymer given 
so far is rather conventional. The
form of  $G_t$ in Eq.~(\ref{P}) describes a generic polymer.
To appreciate the {\it glassy} nature of the pinned phase,
it is necessary to go beyond the  description 
in terms of the mean $\overline{P_1}$.
We first note that an exponent $\zeta > 1/2$ in the pinned phase 
implies that the energy scales involved in the pinned phase are large. 
From the first term in (\ref{H}), we see that 
a displacement of order $t^\zeta$ 
of the free end costs
a minimum elastic energy of order $t^{2\zeta-1}$ 
which grows for long polymers.
Growth of the characteristic energy scale for 
order parameter variation with length
scale in conventional systems implies that the 
system is in an ordered phase governed
by a zero temperature fixed point. Although there is 
no ``order parameter'' in the
directed polymer, the displacement $\vec{\xi}(z)$ play 
a similar role and growth of
the energy scale for variations of $\xi$ with length 
scale --- here $t$ --- implies
that the pinned phase is controlled by a zero temperature 
renormalization group fixed
point whose properties control the scaling of various quantities such as 
$G_t$ in Eq.~(\ref{P}).
By analogy with the ordered phase in conventional 
statistical mechanical systems, we
expect that the large scale properties of the pinned phase
 can be described in terms
of a ``ground (or equilibrium) state'' or ``states'', 
and fluctuations about or 
between these ``states''. Thus
  the configuration of the polymer selected by
``thermal" averaging should be
 the equilibrium ``state" that optimizes the total free
energy for a given realization of the random potential. 
At zero temperature for a finite polymer with one
point fixed at $(\vec{x},t)$, there will be a unique 
preferred path (see Fig.~1(a)),
$\vec{\xi}^*(z)$, which is the ground state, i.e., the state with the lowest
energy.  Here, the term ``state" 
refers to an {\it optimal} path starting from the fixed end
at $(\vec{x},t)$. 
A well-defined thermodynamic limit exists for the state 
of a semi-infinite polymer
if  the thermal mean position $\avg{\vec{\xi}(z)}$ and all
other properties of the polymer at {\it fixed} finite $z/t$  tend to a unique
limit for a specific sample as  $t\to \infty$. 
The conjecture that this holds for
almost all samples was made and supported in Ref.~\cite{fh}.
 The above definition of state
 only makes sense at $T=0$ after providing short distance cutoffs 
$a_\xi$ and $a_\tau$ in the $\vec{\xi}$ and $z$ directions respectively. 
At small but finite temperatures, thermal fluctuations wash out the 
effect of disorder at short length scales, and the polymer does not 
``feel" the random potential until scales $a_\xi(T)$ and $a_\tau(T) 
\sim a_\xi^{1/2}(T)$~\cite{fh}. For example, we have in $d=1$ 
\begin{equation}
a_\xi(T) = T^3/\kappa D.
\end{equation} 
 We can then study the ``states'' of the polymer 
coarse-grained on the scales $a_\xi(T)$ and 
$a_\tau(T)$. A natural conjecture is that at finite temperature
 the equilibrium state
is still unique. What does this mean ? If the 
 one-point distribution function, $P_1(\vec{y},t_0|\vec{x},t)$
for a {\it typical} sample has 
the general behavior sketched in Fig.~1(b) (solid line), i.e.,  
a sharply-peaked function of $\vec{y}$ centered about
some $\vec{y}^*=\vec{\xi}^*(t_0)$, with a width of order $a_\xi(T)$ 
for $\tau = t-t_0 \gg a_\tau(T)$, then this, together 
with similar behavior for all $0\le z \le t$,
 implies that there is a well-defined ``state''
$\vec{\xi}^*(z)$ at long length scales at finite temperature.  As we shall
see, this simple picture is roughly correct but with subtle modifications of
the meaning of ``typical'' which crucially affect the physics.
Note that the equilibrium state is the minimum of the {\it coarse-grained}
Hamiltonian which includes effects of the entropy of small scale fluctuations.
Thus, in general, as shown in Ref.~\cite{fh}, the equilibrium  states for
different temperatures of the {\it same} sample will be very different 
on long length scales.

The peak in $P_1$ (solid line in Fig.~1(b)) should be located within 
a transverse distance $\tau^\zeta$ of   $\vec{x}$ for most samples, 
since the disorder-average  $\overline{P}_1$, sketched as the dashed curve 
in Fig.~1(b), has the characteristic scale 
$|\vec{y}-\vec{x}| \sim \tau^\zeta$.
 The sharpness of $P_1$ for a typical sample
can be revealed by probing disorder moments of 
$P_1$ for {\it different} end points.
For instance, from the joint two-point correlation function,
\eqn{P3}{Q_t(\vec{y}_1-\vec{x},\vec{y}_2-\vec{x},t-t_0) 
\equiv \overline{P_1(\vec{y}_1,t_0|\vec{x},t)P_1(\vec{y}_2,t_0|\vec{x},t)},}
 we can obtain the mean square of the (thermally averaged) displacement,
\begin{equation}
\overline{\avg{ \vec{X}_t(\tau)}^2} = \int d^d\vec{y}_1d^d\vec{y}_2
 (\vec{y}_1-\vec{x})\cdot (\vec{y}_2-\vec{x}) \
Q_t(\vec{y}_1-\vec{x},\vec{y}_2-\vec{x},\tau),
\end{equation}
which is expected to behave as 
\begin{equation}
\overline{\avg{\vec{X}_t(\tau)}^2} 
\approx\tau^{2\zeta}f'(\tau/t), \label{moment2}
\end{equation}
just like $ \overline{\avg{|\vec{X}(t)|^2}}$ in Eq.~(\ref{moment1})
in the pinned phase of the directed polymer but 
with a different scaling function $f'$.
 From Eqs.~(\ref{moment1}) and
(\ref{moment2}), we obtain the mean square of the {\it thermal} fluctuations
about $\langle \vec{\xi}(t)\rangle$,
\begin{eqnarray}
&\overline{C}_T(\tau,t) &\equiv \overline{\avg{ \left| \vec{\xi}(t-\tau) 
- \avg{\vec{\xi}(t-\tau)}\right|^2} }
= \overline{ \avg{|\vec{X}_t(\tau)|^2} }
 - \overline{ \avg{\vec{X}_t(\tau)}^2 } \nonumber\\
& &= \frac{1}{2} \int d^d\vec{y}_1 d^d\vec{y}_2 |\vec{y}_1 -\vec{y}_2|^2
Q_t(\vec{y}_1-\vec{x},\vec{y}_2-\vec{x},\tau).
\end{eqnarray}
One would hope that this quantity $\overline{C}_T(\tau,t)$
 would characterize the typical width of the
one point distribution $P_1$ (solid line of Fig.~1(b)). 
From the above discussion,
the conjectured uniqueness of the equilibrium states at 
finite $T$ would suggest that
$\left[\overline{C}_T(\tau,t)\right]^{1/2}$ would be 
the diameter of the ``tube'' in which the
polymer fluctuates, with $\overline{C}_T(\tau,t)$ not 
growing with $\tau$ or $t$.
 We shall see
however, that in fact $  \overline{C}_T(\tau,t) \sim \tau$ 
in both the pinned and the high
temperature phases.  How is this consistent with the 
existence of an equilibrium state
of the problem in the pinned phase ?  It has been 
argued~\cite{fh} that for {\it typical}
samples (in fact in almost all samples for large $\tau$) 
$C_T(\tau,t)$ is of order unity, but {\it
rare} samples have sufficiently large $C_T(\tau,t)$ that 
they dominate the sample (or
disorder) average.  In this paper, we will see that this 
conclusion arises in a very
natural way and can be characterized analytically by a power law tail in the
distribution of various quantities that are related to $C_T(\tau,t)$.

In conventional critical phenomena, it is often sufficient 
to characterize a system
by characterizing a few moments of the fluctuations. However,
for many random systems,  such averages
do not give an adequate account of what actually goes on in
a {\it typical} sample, since some properties are not  ``self-averaging''. 
The pinned phase of the directed polymer exhibits exactly 
this type of behavior
\cite{fh,m,z} and thus cannot be well characterized by the 
knowledge of a few moments.
For instance, we shall see that a small fraction of samples have nearly 
degenerate ground ``states", $\vec{\xi}^*_a(z)$ and $\vec{\xi}^*_b(z)$, with
segments (say the free ends $\vec{\xi}^*_a(0)=\vec{y}_a^*$ and
$\vec{\xi}^*_b(0)=\vec{y}_b^*$) located far away (i.e., 
$|\vec{y}^*_a - \vec{y}^*_b| \gg a_\xi(T)$), 
as shown in  Fig.~2(a). Such states would be
manifested in $P_1$ as widely separated peaks (Fig.~2(b)) at low temperatures.
The occurrence of such degenerate ``states'' may be rare. However,
if they survive in the thermodynamic limit, with
$|\vec{y}^*_a - \vec{y}^*_b|\to \infty$ as  $t\to\infty$,
then they can still have important physical consequences, 
because these are just the type of states that give rise to 
 large scale fluctuations at low temperatures.
Such almost degenerate states are the low dimensional 
analog of the ``droplets"
\cite{fh,droplet} conjectured to govern much of the low temperature 
properties of spin glasses.
The investigation of these large fluctuations is the focal point 
of this study.  We are particularly interested in obtaining the 
statistics of the almost degenerate states
and understanding the effects of these states 
on physical observables. We shall give a detailed account
of these effects in the following sections.

\subsection{ Distribution Functions}

There are a number of ways one can characterize the rare fluctuations.
 To find the relative abundance of samples
which behave as Figs.~1 and 2, we can, for example,
consider the probability that $P_1$ has the structure shown in Fig.~2,
with two peaks separated by a  displacement $\Delta \equiv |\vec{\Delta}|$
at $z = t-\tau$, i.e., at
a distance $\tau$ from the fixed end. It is useful to consider the
larger peak to correspond to the optimal path and 
the other to correspond to an
excitation with displacement $\vec{\Delta}$ away from the optimal path.  For
the second peak to have a reasonable amplitude, 
the excitation free energy must be of
order $T$ or less, i.e., it is a {\it thermally active excitation}, 
which is analogous to an
active ``droplet'' introduced for spin glasses. 
The probability of an active droplet
excitation of displacement $\vec{\Delta}$ at 
a distance $\tau$ from the fixed end is
obtained from the distribution $Q_t$ as~\cite{twocopies}
\eqn{W2}{W_t(\vec{\Delta},\tau) = \int d^d\vec{y}_1 d^d\vec{y}_2  \ 
\delta^d(\vec{y}_1-\vec{y}_2-\vec{\Delta}) 
Q_t(\vec{y}_1-\vec{x},\vec{y}_2-\vec{x},\tau).}
Like the one-point distribution function $G_t$ discussed 
in Sec.II.A, it will be 
convenient to consider 
\begin{equation}
W(\vec{\Delta},\tau) = \lim_{t/\tau\to \infty} 
W_t(\vec{\Delta},\tau) \label{Winf}
\end{equation}
which, for large $\tau$, is the probability distribution 
of active droplets in the bulk of a
semi-infinite polymer. Note that  the somewhat different
distribution of active droplets at the free end ($z=0$)
\begin{equation}
\widetilde W(\vec{\Delta},t) = W_t(\vec{\Delta},t) \label{Wfin}
\end{equation}
is however the most readily measurable quantity numerically.

In the absence of disorder, the ``bare'' distribution 
$W^{(0)}$ is just a gaussian, 
\begin{equation}
W^{(0)}(\vec{\Delta},\tau) = \left(\frac{\kappa}{4\pi T \tau}\right)^{d/2} 
e^{-\frac{\kappa}{4T \tau}\Delta^2} \label{bareW}
\end{equation}
since $Q_t^{(0)}=P_1^{(0)}\cdot P_1^{(0)} $ independent of $t$.
[Note that $W_t(\vec{\Delta},\tau)$ is the probability 
that the distribution $P_1$ has
weight at two points separated by $\vec{\Delta}$. 
Only if the samples in which this
occurs behave as in Fig.~2 will the interpretation 
of this as the probability of
a  well-defined active excitation really be useful. 
In the absence of randomness,
this interpretation is clearly incorrect.]
With disorder, $Q_t$ becomes highly nontrivial,
and it is not easy in general to compute even the first few
moments of  $W_t$. The large-$\Delta$ tail of the distribution,
 which governs the large scale fluctuations
at low temperature, is difficult to obtain both analytically and numerically.
In Section III, we introduce a free-energy functional which enables us
to compute  the asymptotic form  of $W(\vec{\Delta},\tau)$ explicitly.
We will show that the tail of $W(\vec{\Delta},\tau)$ 
[and hence also the tail of $\widetilde W(\vec{\Delta},t)$ if we ignore the
dependence on $t/\tau$] has  a power law form.

One of the physical consequences of the nearly degenerate ground states is
the behavior of the mean square thermal fluctuations of a given segment 
of the polymer, $C_T(\tau,t) = \avg{\Delta^2}/2$,
which has an average value
\begin{equation}
\overline{C}_T(\tau,t) =
\frac{1}{2}\overline{\Delta^2} = 
\frac{1}{2}\int d^d\Delta \Delta^2 W_t(\vec{\Delta},\tau).
\end{equation}
 This quantity is related to the susceptibility of the polymer
 to a tilt, i.e., the response to 
a term added to the Hamiltonian (\ref{H}),
\eqn{H2}{{\cal H}_{\vec{h}}[\vec{\xi},\eta]
 = {\cal H}[\vec{\xi},\eta] - \int_{t_0}^t dz \ 
\vec{h}\cdot\frac{d\vec{\xi}}{dz}.}
Thermal fluctuations of $\vec{\xi}(t_0)$ at a distance $\tau=t-t_0$ from
the fixed end are simply proportional to the linear
response~\cite{offdiagonal} of the polymer 
by the fluctuation-susceptibility relation, 
\eqn{chi}{\chi[\eta] \equiv 
\frac{\partial}{\partial h_i}{\avg{\xi_i(t)-\xi_i(t_0)}_{\vec{h}\to 0} } 
= \frac{1}{Td} C_T(t-t_0,t),}
where $\xi_i$ is a component of $\vec{\xi}$, and the subscript $\vec{h}$
denotes thermal average taken with respect to the new partition
function, 
\begin{equation}
Z(\vec{x},t;\vec{h}) = 
\int{\cal D}[\vec{\xi}]\ \D{\vec{x}}{t} e^{-{\cal H}_{\vec{h}}/T}.
\end{equation}  
Since the random potential $\eta$ in the Hamiltonian (\ref{H})
 does not single out any
preferred direction in the ($\vec{\xi},z$) space, 
a {\it statistical tilt symmetry} --- related to the
 translational symmetry ---
exists which is recovered upon disorder average.  As a result, one can prove
straightforwardly that on average, the applied tilt $\vec{h}$ merely
shifts the mean position of the polymer at $t_0$ to
$\overline{\avg{\vec{\xi}(t_0)}}_{\vec{h}} 
= \vec{x} -  \vec{h} \tau/\kappa$ (see Appendix
C and Ref.~\cite{Schulz}).
The disorder-averaged susceptibility is thus just
\begin{equation}
\overline{\chi} = \frac{\tau}{\kappa}  
= \frac{1}{2Td} \overline{\Delta^2}, \label{sym}
\end{equation}
which is {\it exactly} the same as the susceptibility of the pure system.

If the statistical properties of the ground
states are complicated as illustrated in Figs.~1(a) and 2(a), then
the mean susceptibility $\overline{\chi}$ does {\it not} provide an adequate
characterization. On the one hand, a sample such 
as the one shown in Fig.~1(a) 
will contribute very little to the average susceptibility at low temperatures
 since it is locked in a unique state, separated from the lowest
excited state by a  free energy difference $\gg T$.
On the other hand, a sample corresponding to
Fig.~2(a) has nearly degenerate states. Thermal fluctuations will thus cause
large scale ``hopping" of the polymer from one state to the other.
Thus the latter samples can give large contributions to 
the average susceptibility
even though they occur rarely. The relative abundance of such
samples is given by the susceptibility distribution function,
\begin{equation}
D_t(\chi,\tau) = \overline{\delta(\chi-\chi[\eta])}.
\end{equation}
A first principles calculation of $D_t$ would require the knowledge of 
the full distribution of the function $P_1$, 
or at least all of its correlations,
$$ \overline{P_1(\vec{y}_1,t_0|\vec{x},t)\cdots
P_1(\vec{y}_n,t_0|\vec{x},t)};
$$
this will not be attempted here. To obtain the
qualitative behavior, we will instead {\it assume} 
that a sample with an active
droplet of size $\Delta$ has a susceptibility of order 
$\chi \sim \Delta^2/(2Td)$.
This assumption obviously will not be very good if there are {\it many}
almost degenerate states that contribute equally to the susceptibility, 
but it should be reasonable if large scale degenerate states 
occur only rarely, as we will demonstrate is the actual case. 
Using this approximation,
 the susceptibility distribution can be simply obtained from the
droplet distribution as
\eqn{dchi}{ D_t(\chi,\tau) = \int \frac{d^d\vec{\Delta}}{\Delta^2/(2Td)} 
\ W_t(\vec{\Delta},\tau) \ u[\chi/(\Delta^2/2Td)],}
where $u(s)$ is a sharply peaked scaling function whose precise shape
 depends on more knowledge of the distribution of $P_1$.

\subsection{Free Energy Variations}

The functions $W_t(\vec{\Delta},\tau)$ and $D_t(\chi,\tau)$ introduced above
describe the 
distribution of the polymer's  {\it equilibrium} fluctuations 
and susceptibilities. But
in reality (experimentally or numerically), equilibration 
of a glassy system is often very difficult~\cite{note1}.
The reason is that the nearly degenerate states
that give rise to large scale, low energy droplet excitations 
are typically separated by
large energy barriers, and therefore have extremely long relaxation times. 
Knowledge of the distribution of the barriers
 is therefore crucial in understanding the {\it dynamic} properties of 
the glassy polymers. 

The details of the dynamics at low temperatures 
is very complex and beyond
the scope of this paper. However, one can get a bound of
the free energy barriers {\it within} the equilibrium
theory. To do so, let us consider the nearly degenerate states shown
in Fig.~2(a). To probe the free energy ``landscape" between $\vec{y}_a^*$ and
$\vec{y}_b^*$, we would like to know 
$F(\vec{x},t;\vec{y},0)$, the free energy of a
polymer $\vec{\xi}(z)$ with {\it both}   ends
fixed at $\vec{\xi}(t)=\vec{x}$ and $\vec{\xi}(0)=\vec{y}$, 
and then vary  $\vec{y}$ in between $\vec{y}_a^*$ 
and $\vec{y}_b^*$ (see Fig.~3).
The maximum, $F^*$, of  $F(\vec{x},t;\vec{y},0)$  in this range  of $\vec{y}$
gives a {\it lower bound}  of the free energy
barrier to move the end $\vec{\xi}(0)$ from  $\vec{y}_a^*$ to $\vec{y}_b^*$,
 since the polymer must pass through all intermediate states with 
$\vec{y}_a^* < \vec{y} < \vec{y}_b^*$, i.e., in a range of width $\Delta$. 
Note that this does not take into
account the {\it additional} barriers the polymer 
may encounter at intermediate
states (see Fig.~4). At this point, it is not clear whether or not the
cumulative effects of such barriers will be much
larger than $\delta F^* = F^* - F(\vec{x},t;\vec{y}_a^*,0)$.

Unfortunately, with two ends fixed, $F$ is difficult to compute
analytically. What {\it can} be obtained readily (see Appendices A and B) 
is the statistics of the free energy of a polymer 
of length $t$ with only {\it one} 
end fixed, which is just
\begin{equation}
 F(\vec{x},t) = - T \log Z(\vec{x},t).
\end{equation}
 As shown by the numerical studies of Kardar and
Zhang~\cite{kz}, two polymers with end points fixed 
at a distance $\Delta\lesssim t^\zeta$
apart will merge a distance $\Delta^{1/\zeta}$ from 
the fixed ends and coincide for the
rest of the way (see Fig.~5). Since the free energy 
difference between two polymers
with different end point positions will only arise 
from the section over which they
differ,  we expect the typical free energy difference,
$F(\vec{x},t;\vec{y},0) - F(\vec{x},t;\vec{y}_a^*,0)$, 
of an intermediate state in Fig.~3 with 
$|\vec{y}-\vec{y}_a^*| \sim O(\Delta)$, 
to scale the same way as the difference $F(\vec{x}+\vec{\Delta},t)
- F(\vec{x},t)$ for  the two polymers in Fig.~5. The latter is 
 characterized by the free energy correlation function
\begin{equation}
C_F(\vec{\Delta},t) = \overline{ [F(\vec{x}+\vec{\Delta},t) 
- F(\vec{x},t)]^2 } \label{corr}
\end{equation}
which is conjectured to scale as 
\begin{equation}
C_F(\vec{\Delta},t) \sim \Delta^{2\alpha}
\end{equation}
for $\Delta \ll t^\zeta$, with the exponent known to be {\it exactly}
$\alpha = 1/2$ in 1+1 dimensions. 
Thus $\sqrt{C_F(\vec{\Delta},t)}$ provides the lower
bound for the barrier of formation of 
an intermediate active droplet, which is the
typical scale for $F^*$.

On the other hand, two polymers with end points 
fixed at $\Delta \gg t^\zeta$
 will not overlap at all and thus behave as 
if they are in independent samples.
In this case,
\begin{equation}
C_F(\vec{\Delta},t) \sim t^{2\theta}
\end{equation}
with $\theta$ being the exponent characterizing 
the sample-to-sample free energy
variations. A simple scaling form, which we shall
 see arises naturally, links the two
limits:
\begin{equation}
C_F(\vec{\Delta},t) = \Delta^{2\alpha} \widetilde{c}(\Delta/t^\zeta)
\end{equation}
yielding the scaling relation 
\begin{equation}
\alpha = \theta/\zeta.
\end{equation}
We will later derive the scaling law 
\begin{equation}
\theta=2\zeta - 1
\end{equation}
which arises simply from the naive contribution 
to the free energy from the elastic
part of the Hamiltonian with displacement of order $t^\zeta$.

\newpage
\section{Statistics of  Rare Fluctuations}

In Section II, we defined a number of 
distribution functions and disorder averaged
 correlations functions which are useful
in probing the glassy nature of the randomly-pinned directed polymer.
To obtain these functions in a systematic way, and to uncover the
interconnections among them, we shall use field theoretic
methods. The disorder average will be replaced
 by an average over a weighting
functional, which describes the probability
distribution of the free energy.  This method is inspired by
the Martin-Siggia-Rose dynamic field theory~\cite{msr} 
developed in the context
 of stochastic dynamics onto which 
the directed polymer can be mapped~\cite{hhf,iv}.
The explicit mapping to the noisy-Burgers equation is derived in Appendix B.
Note however, that
the method may not be  limited  to the directed polymer
and is described in Appendix A for an 
arbitrary dimensional manifold for which the 
mapping to stochastic dynamics cannot be performed.
  In this section, we shall use 
this field theoretic method to obtain the tails of the distribution
functions introduced in Section II.C.
We shall limit our discussions to the semi-infinite polymer ($t\to\infty$) 
problem for which the analysis is the simplest.
  We then interpret the results to
characterize the statistics of the large rare fluctuations.

\subsection{Derivation of the Distribution Functions}
\subsubsection{Formalism}

To obtain the tails of the distribution functions $W_t(\vec{\Delta},\tau)$ and
$D_t(\chi,\tau)$ defined in Section~II.C, we need (see Eq.~(\ref{W2}))
the disorder-averaged
function $Q_t(\vec{y}_1-\vec{x},\vec{y}_2-\vec{x},\tau)$ 
defined in Eq.~(\ref{P3}).
As shown in Appendix A, various distribution functions can be
generated by adding a source term to the Hamiltonian, e.g.,
\begin{equation}
{\cal H} \to {\cal H} + \int_0^t dz \widetilde{J}(\vec{\xi}(z),z),
\end{equation}
and then differentiating the corresponding average free energy,
 $\overline{F}(\vec{x},t;\widetilde{J})$ with respect to $\widetilde{J}$.
The resulting expressions are simple for the semi-infinite 
polymer. For instance, 
\begin{equation}
\frac{\delta\overline{F}(\vec{x},t;\widetilde{J})}
{\delta\widetilde{J}(\vec{y},t_0)} =
\overline{P}_1(\vec{y},t_0|\vec{x},t)=G(\vec{x}-\vec{y},t-t_0),
\end{equation}
 and
\begin{eqnarray}
&G_{2,1}&(\vec{x}-\vec{y}_1,\vec{x}-\vec{y}_2,t-t_0) 
\equiv \frac{\delta^2\overline{F}(\vec{x},t;\widetilde{J})}
{\delta\widetilde{J}(\vec{y}_1,t_0)
\delta\widetilde{J}(\vec{y}_2,t_0)}\nonumber\\
& &= -\frac{1}{T}\left[\overline{\avg{\D{\vec{y}_1}{t_0}
\D{\vec{y}_2}{t_0}}} - Q(\vec{y}_1-\vec{x},\vec{y}_2-\vec{x},t-t_0)
\right] \label{P3c} 
\end{eqnarray}
 In terms of  the noisy-Burgers
equation, the limit $t\to\infty$ corresponds to the statistical steady
state, and $G$ and $G_{2,1}$ are the linear and nonlinear response functions
respectively (see Appendix B).
It is instructive to consider the meaning of $G_{2,1}$ in the limit
of zero temperature (with an appropriate 
short distance cutoff on the random potential):
$G_{2,1}(\vec{x}-\vec{y}_1,\vec{x}-\vec{y}_2,\tau)$ 
is the limit of small $\epsilon$
of $1/\epsilon$ times the probability that the optimal paths passing through
 $\vec{y}_1$ and $\vec{y}_2$ a distance $\tau$ from the fixed end at
($\vec{x},t$) both have energy within $\epsilon$ of the ground state.
Since triple degeneracies are unlikely, this is non-zero when the ground state
is doubly degenerate and $G_{2,1}$ 
is the ``density of states'' of degeneracies.
In terms of $G_{2,1}$, the droplet distribution is
\begin{equation}
W(\vec{\Delta},\tau) = \delta^d(\vec{\Delta}) + W'(\vec{\Delta},\tau),
\end{equation}
with
\eqn{W'}{W'(\vec{\Delta},\tau) = T  \int d^d\vec{y}_1 d^d\vec{y}_2
\delta^d(\vec{y}_1-\vec{y}_2 -\vec{\Delta})
G_{2,1}(\vec{x}-\vec{y}_1,\vec{x}-\vec{y}_2,\tau).}

The free-energy functional described in Appendix A allows us 
to express higher-order
distribution functions such as $G_{2,1}$ in terms of the one-point function 
$G$ (see Eq.~(\ref{G3})). For the directed polymer, the relevant result is
\begin{eqnarray}
\FL &G_{2,1}(\vec{x}-\vec{y}_1,&\vec{x}-\vec{y}_2,t-t_0) = -\int d^d\vec{x}'
d^d\vec{y}_1' d^d\vec{y}_2' dt' dt_1 dt_2
 G(\vec{x}-\vec{x}',\tau-t')\nonumber\\
& &\Gamma_{1,2}(\vec{x}'- \vec{y}'_1, t'-t_1;\vec{x}' - \vec{y}'_2, t'-t_2) 
G(\vec{y}_1-\vec{y}_1',t_1-t_0) G(\vec{y}_2-\vec{y}_2',t_2-t_0),
\label{convol}
\end{eqnarray}
where $\Gamma_{1,2}$ is a ``vertex function'' 
which is natural in the context of 
the noisy-Burgers equation; it will  be specified shortly.
The expression can be represented diagrammatically as in
Fig.~6. We see that the joint distribution $G_{2,1}$ has been
conveniently broken up as a convolution of a number of
 one-body distribution functions (the $G$'s), 
with the branching process controlled
by the vertex $\Gamma_{1,2}$.

For a (statistically) translationally invariant system, it
is convenient (after disorder averaging) to work in  Fourier space, with
\begin{equation}
\widehat{G}(\vec{q},\tau)  = \int d^d\vec{r}\, 
G(\vec{r},\tau) e^{i\vec{q}\cdot\vec{r}}.
\end{equation}
The Fourier transform, $\widehat{W}'$, of
$W'$ in Eq.~(\ref{W'}) becomes
\begin{eqnarray}
&\widehat{W}'(q,t-t_0) = - T &\int^t_0 dt' 
\int^t_{t_0} dt_1 \int^t_{t_0} dt_2
\widehat{G}(0,t-t')\nonumber \\
& &\widehat{\Gamma}_{1,2}(\vec{q},t'-t_1;-\vec{q},t'-t_2) 
\widehat{G}(\vec{q},t_1-t_0)\widehat{G}(-\vec{q},t_2-t_0),
\end{eqnarray}
where $\widehat{\Gamma}_{1,2}$ is the Fourier transform of 
the vertex function $\Gamma_{1,2}$.
 From the scaling form (\ref{Pinfinite}) for $G$, we have 
\eqn{hG}{\widehat{G}(\vec{q},\tau) \approx \widehat{g}(q\tau^\zeta),}
where $q=|\vec{q}|$. Also, $\widehat{G}(0,t)= 1$
 by normalization of the probability
distribution $P_1$. Therefore we have the simple expression
\eqn{hW}{\widehat{W}'(q,\tau) = - T \int^\tau_{-\infty} d\tau' 
\int^\tau_0 d\tau_1 \int^\tau_0 d\tau_2 \ 
\widehat{\Gamma}_{1,2}(\vec{q},\tau'-\tau_1;-\vec{q},\tau'-\tau_2) 
\widehat{g}(q\tau_1^\zeta)\widehat{g}(q\tau_2^\zeta).}

To procede further, we need to know something 
about the vertex $\widehat{\Gamma}_{1,2}$.
We are especially interested in the small $q$ 
behavior since that is what will control
the {\it tail} of the distribution $W(\vec{\Delta},\tau)$.
Note that the normalization of $W$
requires that  $\widehat{W}'(q\to 0, \tau) \to 0$,
hence we must have $\widehat{\Gamma}_{1,2}(q\to 0)\to 0$ 
since  $\widehat{g}(0)=1$.
Also, we recall that the mean susceptibility is  
$\overline{\chi} = \tau/\kappa$
 due to the statistical tilt symmetry (Appendix C), and 
$\overline{\chi} = \frac{1}{2Td} \overline{\Delta^2}$ 
from the fluctuation-susceptibility relation
(\ref{sym}). Thus, 
\begin{equation}
 \lim_{\tau\to \infty}\lim_{q\to 0} 
-\vec{\nabla}_{\vec{q}}^2\widehat{W}'(q,\tau) 
= \frac{1}{d}\overline{\Delta^2} 
=   \frac{2T}{\kappa} \tau.\label{dW2}
\end{equation}
Since $\vec{\nabla}_{\vec{q}}\,\widehat{g} = 0$ by symmetry, 
we must have from Eq.~(\ref{dW2})
\begin{equation}
\widehat{\Gamma}_{1,2}(\vec{q},\tau'-\tau_1;-\vec{q},\tau'-\tau_2) 
= \frac{1}{\kappa} q^2 \gamma(\vec{q},\tau'-\tau_1;-\vec{q},\tau'-\tau_2),
\end{equation}  with
\eqn{gamma}{\lim_{q\to 0}\int^\tau_{-\infty} d\tau' \int^\tau_0 d\tau_1 
\int^\tau_0 d\tau_2 \ 
\gamma(\vec{q},\tau'-\tau_1;-\vec{q},\tau'-\tau_2) = \tau.}

The simplest form of $\widehat{\Gamma}_{1,2}$ satifying Eq.~(\ref{gamma}) is
\begin{equation}
\widehat{\Gamma}^{(0)}_{1,2}(\vec{q},\tau'-\tau_1;-\vec{q},\tau'-\tau_2)
 = \frac{1}{\kappa} q^2 \delta(\tau'-\tau_1) 
\delta(\tau'-\tau_2). \label{bare}
\end{equation}
This is actually the form of the ``bare" vertex, i.e., it is the {\it
exact} vertex in the absence of the 
random potential, as the bare distribution 
$W^{(0)}$ in Eq.~(\ref{bareW}) is readily recovered by substituting
Eq.~(\ref{bare}) in Eq.~(\ref{hW}). In Appendix C, we use
the statistical tilt symmetry to derive some Ward identities
 which ensure that the {\it scaling} behavior of the droplet distribution
obtained from the full vertex 
$\widehat{\Gamma}_{1,2}$ will have the same {\it form}
as that obtained by  using the bare vertex $\widehat{\Gamma}^{(0)}_{1,2}$. 
Using Eq.~(\ref{bare}) then, Eq.~(\ref{hW}) becomes
\begin{equation}
\widehat{W}'(q, \tau) \approx - \frac{T}{\kappa} q^{2} \int_0^\tau d\tau'
\left[\widehat{g}\left(q(\tau')^\zeta\right)\right]^2 \approx
 - \frac{T}{\kappa}
 q^{2-\zeta^{-1}} \widehat{w}(q^{1/\zeta}\tau), \label{cusp}
\end{equation}
where $\widehat{w}(s)$ is a scaling function whose precise form depends on
the actual forms of $\widehat{g}$ 
and $\widehat{\Gamma}_{1,2}$ (see Appendix C), 
but whose limits are simple, i.e.,
$\widehat{w}(s\to\infty) = {\rm const}$ and $\widehat{w}(s\to 0)\to s$.

\subsubsection{Results}

The Fourier transform of $W$ itself, $\widehat{W}=1+\widehat{W}'$,  
has a power-law singularity (an inverted cusp) for small $q$ in the limit
$\tau\to \infty$ as long as $\zeta > 1/2$.
The singularity is only cutoff by $\tau$, 
with $\widehat{W} \approx 1 - (T/\kappa) q^2 \tau$
for $q < \tau^{-\zeta}$. 
Inverse Fourier transforming $\widehat{W}$, we obtain the final result
\eqn{W4}{W(\vec{\Delta},\tau) \approx \frac{1}{\Delta^{d+2-\zeta^{-1}}}
\widetilde{w}(\Delta/\tau^\zeta) 
\qquad {\rm for} \qquad \Delta \gg a_\xi(T),} 
with the scaling function $\widetilde{w}(s)$ having the form 
 $\widetilde{w}(0) = {\rm const}$
and $\widetilde{w}$ rapidly decreasing for $s \gg 1$. 
This result suggests the following  scaling form  for the full distribution
for a polymer of length $t$,
\begin{equation}
W_t(\vec{\Delta},\tau) = \frac{1}{\Delta^{d+2-\zeta^{-1}}}
\widetilde w_{t/\tau}(\Delta/\tau^\zeta), \label{W4full}
\end{equation}
with $\widetilde w = \widetilde{w}_\infty$.
Assuming that $W_t$ has a weak dependence on $t/\tau$ as in $G_t$, 
the above result leads to a power
law distribution of active droplets at the free ends,
\begin{equation}
\widetilde W(\vec{\Delta},t) = \frac{1}{\Delta^{d+2-\zeta^{-1}}}
\widetilde w_1(\Delta/t^\zeta). \label{W4fin}
\end{equation}
which is cut off only by the finite length of the polymer.
A similar result is obtained for the tail of the susceptibility distribution.
From Eqs. (\ref{dchi}) and (\ref{W4full}), we have, 
\eqn{D2}{D_t(\chi,\tau) \approx \frac{1}{\chi^{2-(2\zeta)^{-1}}}
\widetilde{d}_{t/\tau}(\chi/\tau^{2\zeta})\qquad 
{\rm for} \qquad \chi \gg a_\xi^2/Td,} with
 $\widetilde{d}_{t/\tau}$ being 
another scaling function which is qualitatively similar to
$\widetilde{w}_{t/\tau}$.

Note that Eqs.~(\ref{W4}) through (\ref{D2}) 
should only hold in the scaling limit $\Delta \gg a_\xi(T)$,
where the one-point function has the scaling form (\ref{hG}). 
To understand the form of crossover of $W(\vec{\Delta},t)$ from 
$\Delta < a_\xi$ to  $\Delta > a_\xi$, it is useful to
 consider the following approximate form of the one-point function,
\eqn{hG1}{\widehat{G}(\vec{q},\tau) 
= e^{-\frac{T}{2\kappa}\widehat{\nu}(q)q^2 \tau}}
where $\widehat{\nu}(q) = 
1 + (q a_\xi)^{\zeta^{-1} - 2}$.  This simple form of $\widehat{G}$
extrapolates smoothly between the bare function
$\widehat{G}_0(\vec{q},\tau) 
= e^{-\frac{T}{2\kappa} q^2 \tau}$ for $q a_\xi \gg 1$
and  the appropriate scaling form (\ref{hG}) 
for $qa_\xi \ll 1$. It is therefore a useful
guide to the qualitative features of the crossover. (A better determination
of the response function is given in Ref.~\cite{HF}.) 
  Using the bare vertex Eq.~(\ref{bare}) and the approximate form 
Eq.~(\ref{hG1}) for $\widehat{G}$,
the droplet distribution becomes
\begin{equation}
\widehat{W}(\vec{q},t) = 1 - \frac{1}{\widehat{\nu}(q)} 
\left\{ 1 - e^{-\frac{T}{\kappa}\widehat{\nu}(q)q^2 \tau}\right\},
\end{equation}
with the limiting forms
\begin{eqnarray}
\widehat{W}(\vec{q},\tau) = \left\{
\begin{array}{lll}
1 - \frac{T}{\kappa} q^2 \qquad &{\rm for} \qquad & q < \tau^{-\zeta}, \\
- q^{2-\zeta^{-1}} \qquad &{\rm for} 
\qquad & a_\xi^{-1} > q > \tau^{-\zeta}, \\
e^{-\frac{T}{\kappa}q^2 \tau} \qquad &{\rm for} \qquad & q > a_\xi^{-1}.
\end{array}
\right. \label{AsymW}
\end{eqnarray}
Fourier transforming leads to a droplet distribution $W(\vec{\Delta},\tau)$
sketched in Fig.~7 (solid line), 
with the asymptotic scaling behavior for $\Delta \gg
a_\xi$ given by Eq.~(\ref{W4}), and a smooth behavior for $\Delta < a_\xi$. 
It should be emphasized 
that the precise form of $\widehat{G}$ 
and $\widehat{\Gamma}_{1,2}$ will change only the details of 
the crossover function connecting 
the scaling region to the regions with $\Delta <
a_\xi$ and $\Delta > \tau^\zeta$, 
but not the qualitative features of the distribution.
Assuming weak dependence of 
$W_t(\vec{\Delta},\tau)$ on $t/\tau$, we expect the
distribution $\widetilde W(\vec{\Delta},t)$ for the end point to have the
same form (solid line of Fig.~7) with yet another crossover function. Finally,
the same considerations lead to a similar form for the distribution
of the susceptibility, $D_t(\chi,\tau)$. 

\subsection{Glassy Properties of the Pinned Phase}
\subsubsection{Droplet Excitations}

We now interpret the results obtained in the previous sections in terms of 
 the structure of ground states and excitations of 
 the randomly-pinned directed polymer. 
We first compare the result for 
$W(\vec{\Delta},\tau)$ in the limit of zero disorder.
In this case, the microscopic cutoff length diverges, 
i.e., $a_\xi \to \infty$,
and the distribution $W^{(0)}(\vec{\Delta},\tau)$
no longer has a power-law tail and instead 
takes on a simple  gaussian form Eq.~(\ref{bareW}) 
sketched as the dashed line in
Fig.~7. Comparing the solid and the dashed lines, we clearly see
that the biggest effect of the disorder is to shift the distribution to 
the small-$\Delta$ end.  From Fig.~7, it is clear that most of 
the samples have $\Delta \ll \tau^\zeta$. In fact, 
the likely end point separations 
for typical samples is given by the peak of the distribution which occurs for
 $\Delta \lesssim a_\xi$.
This suggests that {\it most} samples have a {\it unique} ground state 
(and a unique equilibrium state at low $T$), like the one
sketched in Fig.~2(a). Only a small fraction of the samples, of the order
\eqn{w}{U_t(\tau) = \int_{\tau^\zeta}^{2 \tau^\zeta}
 d^d\vec{\Delta} \ W_t(\vec{\Delta},\tau) 
\sim \tau^{-\widetilde\theta}u(\tau/t),
\qquad {\rm with} \qquad \widetilde\theta = 2\zeta - 1,}
have almost degenerate ground states that are separated by a distance of
$O(\tau^\zeta)$ at a distance $\tau\gg a_\tau$ from the fixed end. These are
the large scale active droplets excitations 
 depicted in Fig.~2(b). It has been
argued in Ref.~\cite{fh} that $\zeta > 1/2$ for the low temperature phase
of the directed polymer in any finite dimensions. If this is indeed the case, 
then $\widetilde\theta > 0$ and from Eq.~(\ref{w})
 the large droplets are rarely encountered 
in the thermodynamic limit $\tau\to\infty$. However,
the  rare occurrence of these droplets can dominate disorder-averaged
thermodynamic quantities due to the diverging sizes of 
active excitations when they 
do occur. For instance, it is straightforward
to verify that the mean square end point fluctuations,  
$\overline{C}_T(\tau,t) 
= \frac{1}{2} \overline{\Delta^2} \sim \tau$,
as required by the statistical tilt symmetry Eq.~(\ref{sym}), 
is {\it dominated} by
the large droplets in the tail of the distribution 
$W_t(\vec{\Delta},\tau)$, i.e., those
with $\Delta \sim \tau^\zeta$.

At this point, we should reexamine the identification of  
$W_t(\vec{\Delta},\tau)$
 with the droplet distribution: In doing so, we made an implicit assumption 
that configurations 
with multiple almost-degenerate ground states give negligible contributions
 in the thermodynamic
limit. This is clearly wrong in the absence of disorder.
To check the assumption in the presence of disorder, 
we can compute the probability
of having a triply-degenerate ground state, such as the one shown in Fig.~8.
Extending the method described in the above sections, we can consider the 
probability that $P_1(\vec{y},t_0|\vec{x},t)$ has three large peaks at the end
point $t_0 = 0$. (The same can be applied to degeneracies in the bulk.) 
This probability can be obtained from the correlation 
function $\overline{P_1 P_1 P_1}$.
The connected part of this correlation function of
the distribution is formally given
by $G_{3,1} = \delta^3 \overline{F}/\delta \widetilde{J}^3$ 
and can be computed 
from the knowledge
of $G$ and $\Gamma_{1,2}$ (see the Appendices). 
The resulting probability of $\vec{y}_1$,
$\vec{y}_2$, and $\vec{y}_3$ all separated by $O(\Delta)$ 
(Fig.~8), is of order
$\Delta^{2d} W^2(\vec{\Delta},t)$ as one would naively 
expect from the droplet interpretation. 
Therefore, the occurrence of triply-degenerate ground states 
with three end points
all $O(t^\zeta)$ away from each other is very rare, 
of $O(t^{-2\widetilde\theta})$.
One can therefore neglect the effect of such 
multiply-degenerate ground states for most observables
  as long as $\widetilde\theta >0$, i.e., if $\zeta > 1/2$.

The analysis we have carried out yields information, 
as we have seen, about the
probability of finding large thermally active excitations, i.e., those with
excitation free energy $\delta F$ of order $T$ 
or smaller.  In contrast, the natural
quantity that arises in the phenomenological 
scaling theory~\cite{fh} is the probability
$P_D(\Delta, \delta F)$ of finding an excitation 
with displacement of order $\Delta$
and free energy $\delta F\sim \Delta^\alpha$ where $\alpha$ is the exponent
 which controls the magnitude of the free energy variations as the end point
$(\vec{y},0)$
is moved through distances of order $\Delta \ll t^\zeta$ (Sec.II.D). 
Since most of the path will change when $\Delta\sim
t^\zeta$, then $t^{\zeta\alpha} \sim t^\theta$ is the scale of
variations of the free energy (when $\Delta > t^\zeta$), 
and {\it also} the variations
 from sample to sample. The basic scaling hypothesis 
is that all free energy scales
 on length scale $\tau$ scale as
$\tau^\theta$ and all displacement $\Delta$ scale 
as $\tau^\zeta$,  so that for
$a_\xi \ll \Delta \ll t^\zeta$,
\begin{equation}
P_D(\Delta, \delta F) d(\delta F) \sim \frac{d(\delta F)}{(\delta F)^\alpha} 
\widetilde p_D(\delta F/\Delta^\alpha).
\end{equation}

From balancing the change in the {\it mean} 
free energy as a function of $\vec{y}$
(which is just $\kappa |\vec{y}-\vec{x}|^2/(2t)$ 
from the statistical tilt symmetry),
with the random part of the free energy as 
$\vec{y}$ is varied, the exponent identity
\begin{equation}
\theta = 2\zeta - 1 \label{id}
\end{equation}
is obtained. It was argued in Ref.~\cite{fh}
 that the natural form of the scaling
function $\widetilde p_D$ is that it goes to 
a constant for small argument so that for $T
\ll \Delta^\alpha$, the probability of an excitation 
being thermally active is of
order $T/\Delta^\alpha$.  Although ``natural'' 
and consistent with numerical studies,
this does  appear as an additional assumption.  
We now see, however, that the present
calculation tells us that  it is correct since we have found that 
\begin{equation}
P_D(\Delta, \delta F=T) \sim \int_\Delta^{2\Delta} 
W(\Delta',t)d^d\Delta' \sim \Delta^{-2+\zeta^{-1}},
\end{equation} which agrees with the conjectured form. The exponent
$\widetilde\theta$ for active excitations with $\Delta \sim t^\zeta$ is
therefore equal to $\theta$.
We have thus shown that the exponents describing the probability of the {\it
typical} and the {\it rare} active excitations are the same! This result
is rather important, since its analog 
has been assumed in a variety of random systems.

Before discussing the physical response functions, we briefly consider what
would happen if the wandering exponent 
$\zeta$ were $1/2$. It has been suggested, on
the basis of results on the Calyley tree~\cite{cayley}, 
that this might occur for the
pinned phase in sufficiently high dimensions.  
It also occurs at the {\it critical}
 point, separating the high and low temperature phases 
for $d>2$~\cite{doty}.  If $\zeta = 1/2$,
but the displacements at a distance $\tau$ from the fixed end 
grow logarithmically faster than $\tau^{1/2}$, then $\theta$ would
be zero but there will be logarithmic growth 
of the free energy scale with length
scale and, presumably, a $(\log q)^x$ cusp in 
$\widehat{W}(q,\tau)$ which would yield   
a distribution of the form $W(\vec{\Delta},\tau) 
\sim \Delta^{-d}(\log \Delta)^y$ 
in the scaling region. In this case, configurations 
involving multiply degenerate
ground states could contribute substantially 
to disorder averages, and the droplet
picture would no longer be simple. If $\zeta$ 
is strictly $1/2$ with no logarithms,
the behavior would be even more complicated.
Exact calculations on the Cayley tree~\cite{cayley}
 where this occurs shows that,
 indeed a finite fraction of samples
have nearly degenerate ground states with very little overlap between them.
However, many of the properties on the Cayley tree are pathological~\cite{fh}
 because two paths never meet again once they branch.
At the critical
point of the polymer's glass transition for $d>2$, the free energy
scale is of order $T$ because the transition occurs 
at a finite temperature~\cite{doty}.
Depending on the definition of $\theta$, it may or 
may not include logarithms~\cite{fh}.
In any case,   our considerations
based on low free energy excitations  must be modified 
as in Refs.~\cite{fh} and
\cite{hf}. 
The physics of this glass transition is very rich and will be 
left for future study.

\subsubsection{Anomalous Linear Response}

We now analyze the distribution of the linear response $\chi$.
As explained in Sec.III.A, we obtain a power law tail
for $D_t(\chi,\tau)$ from  assuming that $\chi$ is  
dominated by individual large active
droplets of appropriate sizes.  What were ignored were contributions to $\chi$
from multiply-degenerate ground state configurations such as the one sketched 
in Fig.~8.  But we have shown in Sec.III.B.1
 that the probability of the triply degenerate
configuration is of order $t^{-\theta}$ lower than 
that of the doubly degenerate
configurations. (Higher order degeneracy should be 
even rarer.) On the other hand,
contributions to $\chi$ from a given doubly or triply degenerate configuration
are of the same order ($\sim \Delta^2/T$). 
Thus it is reasonable to neglect the 
contribution  to $D_t(\chi,\tau)$ from higher order degeneracies
 in the large $\tau$ limit. The resulting distribution, which
depends solely on the droplet excitations,  should then be similar in form to
$W_t(\Delta\sim\sqrt{\chi},\tau)$ as sketched in Fig.~7 
(solid line), with a regular part for
small $\chi$ and a long tail described by Eq.~(\ref{D2}) for 
$\chi \gg  (a_\tau/\kappa)^2$.
[Again, this would not hold at the critical point or in 
$d=\infty$ where $\zeta = 1/2$
and $\theta = 0$. In these situations, the 
multiply-degenerate ground states could contribute
to $\chi$ at the same order, and the distribution 
function $D_t(\chi,\tau)$ might 
be more complicated. It will be interesting
to explore these special situations, for example  
by analyzing the directed polymer on a Cayley 
tree, or by studying the behavior {\it at} the glass transition.]

The distribution $D_t(\chi,\tau)$,
and in particular, the distribution of the linear response to a uniform field
\begin{equation}
\widetilde{D}(\chi,t) = D_t(\chi,t)
\end{equation}
 can be obtained experimentally, by following the domain wall of a planar
random Ising magnet in the ordered phase, 
or by following the end points of flux lines
in type-II superconductors close to $H_{c1}$. This may not seem practical 
at first glance, as many samples would seem to be required.
However, thanks to the special statistical tilt symmetry present in this 
problem (see also Ref.~\cite{hf}),
 the distribution $\widetilde D(\chi,t)$ can actually 
be obtained from a {\it single} sample. 
To see this, we just need to realize that
the directed polymer only occupies a very limited volume of a given sample.
To be more specific, the polymer of length $t$ 
is essentially contained within a $d+1$ dimensional
``cone'' (actually, a distorted paraboloid) 
of length $t$ and radius $t^\zeta$ about the pinned end.
So if we apply a uniform tilt field $h$ 
[Eq.~(\ref{H2}) with $t_0=0$], the polymer will
be biased, on average, to tilt by an angle $\phi = h/\kappa$. It is 
therefore forced out of the cone for 
\eqn{hc}{  h > O(t^{\zeta-1}).} Once outside of the cone, 
the polymer encounters a random potential completely different from the one
within the cone, since the random potential inside and outside of the cone are
 uncorrelated. Due to
the statistical tilt symmetry, the random part of 
the new potential the polymer encounters
 is statistically the same as the one inside the cone. In other
words, the new potential is just like another realization of randomness or
another sample except for the trivial mean tilt energy.  
Therefore, disorder averages can be obtained simply by
monitoring the linear response $\chi$ as 
a function of the tilt angle which is varied
by changing the applied field $h$ over a range $\delta h \gg t^{\zeta-1}$.

Recently, M\'{e}zard~\cite{m} reported such a numerical study
of the polymer's response as a function of the applied tilt field $h$,
 for a fixed realization of the random potential in 1+1 dimensions. 
Qualitatively,  the position of the free end $y(h)=\avg{\xi(0)}_h$ is  
found to be {\it independent} of $h$ for 
a range of the applied field. But $y(h)$
can ``jump", by a distance of $O(t^{2/3})$ for a polymer of length $t$, if
the applied field is increased by $O(t^{-1/3})$. As a result, the response
$\chi(h) = dy(h)/dh$ exhibits a series of sporadic
 sharp peaks, with strengths of $O(t^{4/3})$,
separated by distances of $O(t^{-1/3})$, yielding $\bar{\chi} \sim t$
on average. (Here the average is performed over the applied field,
i.e.,  by $ \frac{1}{h} \int_0^h dh' \chi(h')$
for $h \gg t^{-1/3}$.)

The above  qualitative findings are of course consistent with the droplet
picture described in this paper. 
As expected, a directed polymer in a typical realization
of the random potential (or at some given 
tilt field) has an unique ground state.
Typical excited states with distances $O(t^\zeta)$ away are of $O(t^\theta)$
higher in free energy. This ``energy gap" prohibits any large scale response
to a very small increase in the field $h$. 
Since the applied field supplies an energy 
$\vec{h}\cdot (\vec{\xi}(t)-\vec{\xi}(0)) \sim h t^\zeta$, 
the ``energy gap" can be overcome for $ h t^\zeta > t^\theta$. 
This is just the condition Eq.~(\ref{hc}) using the exponent relation
Eq.~(\ref{id}).
In 1+1 dimensions, we have $\zeta = 2/3$. Thus the energy gap is overcome 
and the end point jumps when $h\sim t^{-1/3}$.
By the statistical tilt symmetry,
the process repeats itself (statistically) once the polymer's end point
``jumps" by $O(t^\zeta)$. This then produces  the sporadic jumps in $y(h)$
[or  peaks in $\chi(h)$] observed in Ref.~\cite{m}.  

One might wonder why there are not many jumps 
of size $\Delta \ll t^\zeta$, involving
just a section of length $\Delta^{1/\zeta}$ near
 the free end.  These typically
cost energy of order $\Delta^\alpha$ and gain 
field energy of order $\delta h \Delta$.
Thus naively, they will occur when $\delta h  \sim 1/\Delta^{1-\alpha}$.  
However, since this
increases with decreasing $\Delta$, small jumps
 will almost alway be preempted by
large jumps. They will only occur if the excitation 
energy is anomalously small [of
order $\Delta/t^{1-\zeta}$ which occurs with 
probability $(\Delta/t^\zeta)^{1-\alpha}$].
Therefore most of the end point motion as a function of $h$
 will occur in jumps of $O(t^\zeta)$.

It should be possible to get some analytical 
information on the distribution of
{\it jumps}
from the methods of the paper, but the  details 
are complicated. One would need
information on joint distribution of end points 
with slightly different tilt fields 
$\vec{h}_1$ and $\vec{h}_2$, but the same random potential, such as 
\begin{equation}
\overline{P_1(\vec{y}_1,0|\vec{x},t;\vec{h}_1)
P_1(\vec{y}_2,0|\vec{x},t;\vec{h}_2)}.
\end{equation}

Conceptually, studying the response $\chi$ of a system 
by slowly varying an uniform bias $\vec{h}$
is similar to recent studies of self-organized criticality
 in cellular automaton
models of sandpiles~\cite{btw,kadanoff} and earthquakes~\cite{quake}: 
Tilting the polymer corresponds to tilting the sandpile or increasing
 the loading force on a fault in earthquake models. 
Anomalously large responses
(the sporadic peaks in $\chi(h)$) exhibited by the polymers are like
the ``avalanches'' encountered in the cellular automaton models.
Of course, there are fundamental differences: The present study
addresses {\it equilibrium} fluctuations of 
a disordered system at low temperatures,
while the cellular automaton models mimic the {\it nonequilibrium} dynamics
at zero temperature. 
In particular, if a directed polymer in a random 
potential with relaxational dynamics
at zero temperature is driven by increasing 
the tilt bias adiabatically, the spectrum
of avalanches, i.e., jumps, will be very 
different from the equilibrium behavior
discussed above: There will be many small avalanches 
and only occasionally large ones,
as analyzed for higher dimensional driven manifolds 
by Narayan and Fisher~\cite{nf}. 
In addition, as $h$ is increased, the polymer will stay in configurations 
that for a while become more and more strained with the free end advancing 
much faster than the midpoint, in contrast to the equilibrium case.
Nevertheless, the two problems share the key feature
of a broad distribution of large rare ``events".
Intuition gained from studying the equilibrium
behavior may therefore provide useful hints 
for understanding the nonequilibrium problem.

\newpage
\section{Discussion}

In this paper, we have presented a systematic 
investigation of some properties of
 the  large scale, low energy
excitations of directed polymers in random media. The large scale 
thermally active  excitations occur in 
a small fraction of samples (or regions of a sample) which have
almost degenerate ground states that differ by 
large displacement in the region. 
We showed the probability
of occurrence of such samples, with excitations 
on scale $t$, is of order  $t^{-\theta}$ in the limit 
the polymer length $t\to\infty$. Nevertheless, many of the disorder-averaged
 thermodynamic properties are
dominated entirely by these rare samples.
 This is because in most of the samples, a polymer is essentially frozen
in a unique ground state,  with the  largest scale
 excitations having energy of order
$t^\theta$ hence not being thermally active.
 The rare samples with large active excitations make up
for their small numbers by displaying anomalously  large fluctuations
which dominate the mean. 
We computed the distribution of the almost degenerate ground states 
for a polymer with one end fixed.
While the distribution is peaked at small  separations,
 we found a long power-law tail describing almost 
degenerate configurations that are 
far apart, with  separation $\Delta$ at a distance $\tau$ from the fixed end 
 ranging from zero to the
maximum important extent $\tau^{\zeta}$ (with rapid decay for $\Delta \gg
\tau^\zeta$).
This distribution of almost degenerate configurations also
 provides information on the distribution of excitations on scales
much {\it less} than the system size which dominate 
many of the properties of large systems.

The analysis was carried out by 
exploiting the known mapping of the directed polymer to the noisy Burgers
equation, which converted singular, rare fluctuations to smooth, hydrodynamic 
fluctuations. We found that the interesting  polymer 
distribution functions corresponded  to various
response functions of the Burgers' equation, and 
the tails of the distributions,
which are generally difficult to obtain, are in 
this case just the hydrodynamic
limits of the response functions and therefore 
much more  straightforward to analyze.
In the context of the Burgers' equation, the 
power law tails are merely
the ``long time tails" that are well known in 
hydrodynamic systems. Most of the
difficulties with analyzing the properties of 
the pinned phase of the directed polymer
by conventional methods are due to its control by 
a non-trivial randomness-dominated
zero-temperature fixed point at which temperature 
is ``dangerously irrelevant", meaning
that interesting correlation functions {\it cannot}
 be obtained by setting $T$ to
zero.  The main advantage of the approach via the 
noisy-Burgers' equation, is that the
problem is converted into a much more conventional 
one in which fluctuations play the
dominant role and field theoretic methods can be used. 
 The physics of minimization of
the coarse grained free energy to find ``ground states'' 
and then treatment of the
anomalous rare fluctuations around the ground states 
miraculously appear as a
consequence of the existence of a conventional 
fixed point of the hydrodynamic problem.

Our findings support the main conjectures of the droplet 
theory~\cite{fh,droplet}. What we have demonstrated in this paper is
a formal way of obtaining the statistics of the active droplets and 
various direct physical consequences of their distributions.  
As should be clear from the derivation 
of the distribution functions,
a crucial ingredient  which leads to power-law tails of distributions 
is the existence of the  statistical tilt symmetry.
 It is thus tempting to attribute the scale-invariant distributions
to some analog of the Goldstone modes associated 
with the  tilt symmetry for the {\it ensemble} of 
disordered systems which is ``spontaneously-broken"
 by the selection of the configuration for each specific sample.
It would be important to understand whether in general  such a symmetry is
sufficient to ensure the existence of a power-law tail of large rare
fluctuations in disordered systems controlled by randomness-dominated
 zero-temperature fixed points.

The large rare thermally active excitations in the 
pinned phase of random directed
polymers are closely analogous to the droplet 
excitations proposed to describe the
low temperature phase of Ising spin glasses.  
Their existence in spin glasses
 was argued to depend, as
in the present case, on the existence of a non-trivial 
zero temperature fixed point
which is {\it stable} to thermal fluctuations~\cite{sg}.  
This stability to fluctuations --- and
thus the whole picture --- occur only when the free energy 
exponent $\theta$ is
positive (or perhaps zero with logarithms). 
But what is the analog of the statistical
tilt symmetry for spin glasses ?  In the Edwards-Anderson model, there is a
statistical local gauge symmetry associated 
with changing the sign of one spin and all
the exchange couplings involving that spin.  
This was used, implicitly, to argue for
the existence of large rare thermally active excitations in 
spin glasses~\cite{sg}. It  plays a
role analogous to the statistical tilt symmetry 
of the directed polymer.  Note however
that the statistical gauge symmetry is not exact 
in more realistic models of spin
glasses in which ferro and antiferromagnetic 
exchanges are not equally likely. 
Nevertheless, it is believed that in the spin glass phase,
 the deviations from the
exact symmetry are irrelevant on long length scales.  
A similar phenomenon occurs for
directed polymers on a {\it lattice} in which there 
are short-distance correlations
in the random potential along the polymer.  In this case, 
the statistical tilt symmetry is
{\it not} exact but numerical evidence support 
the conjecture that the deviations from
it are irrelevant at long length scales.

It is finally useful to compare this study with 
a recent uncontrolled field-theoretic
analysis using replicas. In principle, distribution
functions such as $W_t(\vec{\Delta},\tau)$ can be obtained using
the replica method~\cite{mp}. However, difficulties arise in trying to take
 the $n\to 0$ limit which  is not uniquely 
defined from the positive integer $n$.
M\'ezard and Parisi have proposed a gaussian variational ansatz
~\cite{mp} to treat the directed polymer problem. Within this framework,
 it is easy to see that replica-symmetry 
must be broken in order to
recover the rare fluctuations. In such a treatment, the branching 
tree structure of ground states as  a function of their end points
 in physical space (Fig.~6)
 is replaced by an abstract hierarchical ``family'' tree structure 
in replica space. The key to
the power-law tails of  the distribution functions is the overlap
of ``distant family members'', the $u\to 0$ limit of the function 
$[\sigma](u)$ in the notation of Ref.~\cite{mp}.  Within the gaussian ansatz, 
a nontrivial tail is obtained if $[\sigma](u\to 0)\to u^{2/\theta}$.
Thus the replica-symmetry breaking scheme seems
 to be an approximate way of obtaining 
rare fluctuations within the replica formalism~\cite{note2}.
As such, it {\it may} be a useful tool for other problems.
However, the variational approximation is
uncontrolled and the physics enters in 
a very roundabout manner.  It is thus not at
all clear why this approach has 
any advantages over a direct phenomenological scaling
approach as in Ref.~\cite{fh},   renormalization group 
calculations which are exact
on the hierarchical lattice~\cite{migdal}, 
or approximate ``functional" renormalization
group studies~\cite{manifold,ledou}, all of which have the real virtue
 that the physics appears
directly. Furthermore, it is not clear 
if {\it any} real physics is associated with the
notion of ``replica-symmetry breaking''~\cite{note2}. 
As our study implies, the power-law tail of
the rare fluctuations are a rather general consequence of 
the statistical symmetry of the problem; the ill-defined 
concept of ``replica-symmetry breaking"
 is of course not needed at all in our analysis~\cite{cavity} !
However, the simple structure of the 
free energy functional via the noisy-Burgers'
equation, relies {\it strongly} on the 
one-dimensional nature of the directed polymer.
It will be very interesting to see 
if our method can be extended, using the formalism
of Appendix A, to study
higher (but finite) dimensional systems such as  oriented
manifolds in random media~\cite{manifold}.

It may be  comforting to some to find 
that the important physics of the ground state, 
--- in particular power-law distributions 
of  rare, anomalously large fluctuations ---
can be retrieved both by analysis in physical and replica space.  
For the directed polymer, at least, there are no substantial
disagreement between these two approaches. It is possible that controversies
regarding the nature of the low-temperature phase of finite-dimensional
 spin-glasses may have a similar resolution, 
with the many thermodynamic ``states"
of the infinite-range Sherington-Kirkpatrick 
model becoming, in finite dimensions,
just the large scale, low free energy droplet 
excitations of the phenomenological
scaling approach.

\newpage
\acknowledgements
We are grateful to many helpful discussions 
with E. Frey, V. L'vov, and T. Nattermann, and also thank 
 P. W. Anderson and G. Parisi for 
insightful comments. TH acknowledges the hospitality of NORDITA
where part of this work was completed.
This work was supported in
part by the National Science Foundation through Grants Nos. DMR-91-06237,
 DMR-91-15491 and the Harvard University Materials Research Laboratory.

\vfill

\newpage
\appendix
\section{A FIELD THEORY FOR RANDOM SYSTEMS}

In this Appendix, we consider a general field theoretic description
of manifolds in random media. In particular, 
we study a disordered system described 
by a Hamiltonian ${\cal H}[\vec{\xi}(\underline{z}),\eta]$ which is
 defined on the $D$-dimensional manifold $\underline{z}\in\Re^D$, 
with $\vec{\xi}\in\Re^d$ a $d$-component ``order parameter", 
and $\eta(\vec{\xi},\underline{z})$ a quenched
random variable (in our case the potential)  distributed according to
some distribution $p[\eta]$. The one-point restricted partition function is 
\begin{equation}
Z(\vec{x},\underline{t};\eta) = \int {\cal D}\left[\vec{\xi}\right]
\D{\vec{x}}{\underline{t}} e^{-{\cal H}[\vec{\xi},\eta]/T}.
\end{equation}
Various distribution functions can be obtained by adding a source term
\begin{equation}
{\cal H} \to {\cal H} + \int d^D\underline{z} 
\ \widetilde{J} (\vec{\xi},\underline{z}),
\end{equation}
and then differentiating the free energy,
\begin{equation}
F(\vec{x},\underline{t};\widetilde{J};\eta)
= -T \log Z(\vec{x},\underline{t};\widetilde{J};\eta).
\end{equation}
For instance, 
\eqn{G11}
{\frac{\delta\overline{F}(\vec{x},\underline{t};\widetilde{J})}
{\delta\widetilde{J}(\vec{x}_1,\underline{t}_1)}=
\overline{\avg{\D{\vec{x}_1}{\underline{t}_1}} },}
which is the disorder-averaged conditional 
probability of $\vec{\xi}(\underline{t}_1) 
= \vec{x}_1$
given $\vec{\xi}(\underline{t}) = \vec{x}$, and
\eqn{G21}
{\frac{\delta^2\overline{F}(\vec{x},\underline{t};\widetilde{J})}
{\delta\widetilde{J}(\vec{x}_1,\underline{t}_1)
\delta\widetilde{J}(\vec{x}_2,\underline{t}_2)} =
-\frac{1}{T}\overline{\avg{\D{\vec{x}_1}{\underline{t}_1}
\D{\vec{x}_2}{\underline{t}_2}}^{(c)}},}
where the notations for thermal and disorder averages are the same as
those used in Sec.II.A, and the superscript
 $(c)$ denotes the connected (or truncated)
correlations. We will generally be interested in the n$^{\rm th}$ order
 distribution and correlation functions,
\begin{equation}
{\cal G}^{(c)}_{n,1}({\bf x}_1;\dots;{\bf x}_n|{\bf x}) 
= \frac{\delta^n \overline{F}({\bf x};\widetilde{J})}
{\delta\widetilde{J}({\bf x}_1)\dots\delta\widetilde{J}({\bf x}_n)},
\end{equation}
and 
\begin{equation}
{\cal G}_{0,n}^{(c)}({\bf x}_1;\dots;{\bf x}_n) = 
\overline{F({\bf x}_1)\dots F({\bf x}_n)},
\end{equation}
where we have used the shorthand ${\bf x}_i = (\vec{x}_i,\underline{t}_i)$.
Clearly, translational symmetry in space ($\vec{x}$'s) is restored upon
disorder average. In addition, if we consider a very large manifold where the
boundary of $\underline{z}$ is far from 
any of the distances $\underline{t}_i -
\underline{t}_0$, then we also have translational symmetry in the internal
coordinates. For a semi-infinite polymer, this leads to 
\linebreak
${\cal G}_{1,1}^{(c)}(\vec{y},t-t_1|\vec{x},t)=G(\vec{x}-\vec{y},t-t_1)$, 
which is just the one-point distribution function
$\overline{P}_1(\vec{y},t_1|\vec{x},t)$. Also,
 ${\cal G}^{(c)}_{0,2}(\vec{x},t;\vec{y},t)$ is related to
 the free-energy correlation function $C_F(\vec{x}-\vec{y},t)$ in 
Eq.~(\ref{corr}), 
and ${\cal G}^{(c)}_{2,1}(\vec{y}_1,t_1;
\vec{y}_2,t_1|\vec{x},t)=G_{2,1}(\vec{x}-\vec{y}_1,\vec{x}-\vec{y}_2,t-t_1)$
gives the droplet distribution through Eq.~(\ref{W'}).

To carry out the average over disorder, it is advantageous to eliminate the
random variables $\eta({\bf x})$ in favor of a probability density for 
the field $F({\bf x})$ itself, i.e.,
\begin{equation}
 \widehat{\Xi}[F,\widetilde{J}] = 
\int{\cal D}[\eta] \delta(F+ T\log Z[\widetilde{J};\eta])p[\eta],
\end{equation}
so that the mean free energy is
\begin{equation}
\overline{F}({\bf x};\widetilde{J}) 
= \int{\cal D}[F] F({\bf x})\widehat{\Xi}[F,\widetilde{J}].
\end{equation}
It will be more convenient to write $\widehat{\Xi}$ in terms of 
its (functional) Fourier transform $\Xi[\widetilde{F},F]$ as
\eqn{Xi}{ \widehat{\Xi}[F,\widetilde{J}] 
= \int{\cal D}[i\widetilde{F}] 
\,\Xi[\widetilde{F},F] e^{\int d{\bf x}
\widetilde{J}({\bf x})\widetilde{F}({\bf x})}.}
Note that Eq.~(\ref{Xi}) is  similar to 
the dynamic generating functional of Martin, Siggia
and Rose~\cite{msr,janssen} with $\widetilde{F}$ being the ``response field".
For a generic random system, 
the free energy functional $\Xi[\widetilde{F},F]$ will in
general be rather complicated. However, the one-dimensional nature of the
directed polymer can be exploited~\cite{hhf,iv} to derive a functional that is
{\it local} in $F$ and $\widetilde{F}$.
The explicit form of $\Xi[\widetilde{F},F]$   is shown in Appendix B.

After adding a source term to Eq.~(\ref{Xi}) to generate correlations of $F$, 
we obtain the generating functional 
\eqn{cZ}{{\cal Z}[J,\widetilde{J}] 
= \int{\cal D}[i\widetilde{F}]{\cal D}[F]\, \Xi[\widetilde{F},F]\,
\exp\left[\int d{\bf x} \left( \widetilde{J}({\bf x}) \widetilde{F}({\bf x}) +
J({\bf x}) F({\bf x})  \right)\right].}
The distribution and correlation functions 
of interest can now be obtained from
functional derivatives of 
$\Phi[\widetilde{J},J] \equiv \log{\cal Z}[\widetilde{J},J]$, e.g.,
\begin{eqnarray*}
{\cal G}^{(c)}_{m,n}({\bf y}_1;\dots;{\bf y}_m|{\bf x}_1;\dots;{\bf x}_n) &= 
&\frac{\delta^{m+n} \Phi[\widetilde{J},J]}
{\delta\widetilde{J}({\bf y}_1)\dots\delta\widetilde{J}({\bf y}_m)
\delta J({\bf x}_1)\dots\delta J({\bf x}_n)}\\
&= &\overline{\widetilde{F}({\bf y}_1)
\dots\widetilde{F}({\bf y}_m) F({\bf x}_1)\dots F({\bf x}_n) }^{(c)},
\end{eqnarray*}
where the disorder average (overbar) 
is now performed using ${\cal Z}[J=0,\widetilde{J}=0]$.

Note that the normalization of 
$\widehat{\Xi}[F,\widetilde{J}]$ requires ${\cal Z}[J=0,\widetilde{J}] = 1$.
Hence all of the moments ${\cal G}^{(c)}_{m,0} = 0$.
To obtain the other moments, we introduce the vertex functions~\cite{amit},
\eqn{vertex}{\gamma_{m,n}({\bf y}_1;\dots;
{\bf y}_m|{\bf x}_1;\dots;{\bf x}_n) = 
\frac{\delta^{m+n} \Gamma[{\overline{\widetilde{F}}},\overline{F}]} 
{\delta{\overline{\widetilde{F}}}({\bf y}_1)
\dots\delta{\overline{\widetilde{F}}}({\bf y}_m)
\delta \overline{F}({\bf x}_1)\dots\delta \overline{F}({\bf x}_n)},}
where $\Gamma[{\overline{\widetilde{F}}},
\overline{F}]$ is the Legendre transform of $\Phi$:
\begin{equation}
\Gamma[{\overline{\widetilde{F}}},
\overline{F}] = \int d{\bf x}
\left[ {\overline{\widetilde{F}}}({\bf x})\widetilde{J}({\bf x}) 
+ \overline{F}({\bf x})J({\bf x})\right] - \Phi[\widetilde{J},J],
\end{equation}
with $\widetilde{J}$ and $J$  eliminated from the above expression using
\begin{equation}
 {\overline{\widetilde{F}}} = 
\frac{\delta\Phi}{\delta \widetilde{J}},\qquad\qquad
  \overline{F} = \frac{\delta\Phi}{\delta J}.
\end{equation}
It is a simple exercise to show that all of the correlation functions
${\cal G}^{(c)}_{m,n}$ can be 
obtained from the vertex functions $\gamma_{m,n}$~\cite{amit}.
The advantage of introducing the vertex functions
is to allow the derivation of Ward identities from symmetries~\cite{amit}.
This will be done in Appendix C.

With ${\cal G}^{(c)}_{m,0} = 0$, the expressions for 
${\cal G}^{(c)}_{m,1}$ are particularly simple: 
Only the connected parts survive 
(we will therefore drop the superscript $(c)$),
and all of the higher order functions ${\cal G}_{m,1}$
can be obtained from the combination of $\gamma_{1,m}$'s and ${\cal G}_{1,1}$.
For instance,
\eqn{cG11}{ \int d{\bf x} \ {\cal G}_{1,1}({\bf y}|{\bf x}) 
\gamma_{1,1}({\bf x}|{\bf y}') = \delta({\bf y}-{\bf y}'),}
thus $\gamma_{1,1}$ is the inverse ``propagator'', and
\eqn{G3}{{\cal G}_{2,1}({\bf y}_1;{\bf y}_2|{\bf x}) 
= - \int d{\bf x}' d{\bf y}'_1 d{\bf y}'_2
{\cal G}_{1,1}({\bf x}'|{\bf x})\gamma_{1,2}({\bf x}'|{\bf y}'_1;{\bf y}'_2)
{\cal G}_{1,1}({\bf y}_1|{\bf y}_1') {\cal G}_{1,1}({\bf y}_2|{\bf y}_2').}
This leads to Eq.~(\ref{convol}) for the directed polymer with
 ${\cal G}_{1,1}({\bf x}'|{\bf x}) = G({\bf x}-{\bf x}')$, 
$\ {\cal G}_{2,1}({\bf y}_1;{\bf y}_2|{\bf x}) 
= G_{2,1}(\vec{x}-\vec{y}_1,\vec{x}-\vec{y}_2,\tau)$, and
$\gamma_{1,2}({\bf x}'|{\bf y}'_1;{\bf y}'_2) 
= \Gamma_{1,2}(\vec{x}'-\vec{y}_1',t'-t_1';\vec{x}'-\vec{y}_2',t'-t_2')$
due to the translational symmetry in a semi-infinite polymer
 after averaging over the  disorder.
  The result Eq.~(\ref{G3}) is conveniently 
represented by a tree diagram (Fig.~6). 
A higher order distribution function ${\cal G}_{m,1}$ typically contains a
tree diagram involving only $\gamma_{1,2}$ and ${\cal G}_{1,1}$, plus
diagrams involving higher order vertex functions $\gamma_{1,m'}$, 
with $2 < m' \le m$. 
An even larger class of vertex functions
are needed to generate the correlation functions,
${\cal G}_{0,n}$'s. The description of these is more complicated, and we will
not discuss them except to note  that
 for the directed polymer in $1+1$ dimensions,
there exists identities (fluctuation-dissipation theorems)
 linking ${\cal G}_{0,n}$ and ${\cal G}_{n,1}$~\cite{HF,krug}.

\newpage
\section{Mapping to Stochastic Dynamics}

In this appendix, we describe a mapping of the random directed polymer to 
a stochastic dynamics problem --- the noisy Burgers' equation --- 
first noted in Ref.~\cite{hhf}. 
We show how this mapping
can be used to extract information about the directed polymer, 
and thereby provide a concrete example of the general
field theoretic method described in Appendix A.

Let us consider the doubly-restricted partition function
\eqn{Z2}{Z_2(\vec{x},t;\vec{x}',0) = \int{\cal D}[\vec{\xi}] \, \D{\vec{x}}{t}
\D{\vec{x}'}{0} e^{-{\cal H}/T}}
where $(\vec{x},t)$ and $(\vec{x}',0)$ are 
 the end point locations of a directed polymer of length $t$. We see that
$Z_2$ is the Boltzmann weight of 
``propagating'' the polymer from $(\vec{x}',0)$
to $(\vec{x},t)$; it satisfies the ``diffusion" equation
\eqn{diffusion}{ T \frac{\partial}{\partial t} Z_2(\vec{x},t;\vec{x}',0) 
= \frac{T^2}{2\kappa}\vec{\nabla}^2_{\vec{x}} 
Z_2 - \eta(\vec{x},t) Z_2, }
where $t$ is interpreted as ``time" 
and $(\vec{x}',0)$ is merely a parameter, which
enters through the initial condition 
$Z_2(x,t=0;x',0) = \delta^d(\vec{x}-\vec{x}')$.
Since Eq.~(\ref{diffusion}) is linear, 
we can easily integrate over the coordinate
$\vec{x}'$ to obtain the same diffusion equation for $Z(\vec{x},t) \equiv \int
d^d\vec{x} Z_2(\vec{x},t;\vec{x}',0)$, but now with the initial condition
$Z(\vec{x},0) = 1$. Then the free energy of the one-point restricted polymer,
$F(\vec{x},t) = -T \log Z(\vec{x},t)$ satisfies
the well-known noise-driven Burgers' equation~\cite{FNS,KPZ},
\eqn{kpz}{\frac{\partial F}{\partial t} = \frac{T}{2\kappa} \vec{\nabla}^2 F
- \frac{1}{2\kappa} (\vec{\nabla} F)^2 + \eta(\vec{x},t),}
with the initial condition $F(\vec{x},0) = 0$.
 Note that the path integral description of the directed polymer,
i.e., Eqs.~(\ref{H}) and (\ref{Z}), is
very different in form, from the dynamic 
description Eq.~(\ref{kpz})~\cite{KPZ}. 
The task of global optimization presented in the
original directed polymer problem is 
turned into a step-by-step recursive process:
Eq.~(\ref{kpz}) relates the free energy of a polymer of length $t$ to one of 
length $t+dt$ and is thus a transfer 
matrix description of the problem~\cite{cavity}. 

The noisy-Burgers equation Eq.~(\ref{kpz}) 
itself has been the focus of several
 recent studies~\cite{krug}. However, 
despite much effort, no systematic solution 
has been found. The
lone exception is in 1+1 dimensions, where Eq.~(\ref{kpz}) can be thought of 
as the $d=1$ limit of an anisotropic driven diffusion equation~\cite{bks,js}
\begin{eqnarray}
& & \partial_t \phi = \frac{T}{2\kappa} \nabla^2\phi 
- \frac{1}{2\kappa}\partial_x(\phi)^2 +
\eta'(\vec{r},t),\nonumber \\
&{\rm with}  \qquad & \avg{\eta'(\vec{r},t)\eta'(\vec{r}',t')} = -2D \nabla^2
\delta^d(\vec{r}-\vec{r}')\delta(t-t'), \label{dds}
\end{eqnarray}
where $x$ is one particular component of $\vec{r}$. Eq.~(\ref{dds})
 is equivalent to Eq.~(\ref{kpz})
in $d=1$ with the identification $\phi = \partial F/\partial x$ 
and $\eta' = \partial \eta/\partial x$.  
[See Ref.~\cite{ddl} for a detailed discussion 
of the connections between
Eq.~(\ref{kpz}) and Eq.~(\ref{dds}).]
Eq.~(\ref{dds}) has also been investigated in detail~\cite{bks,js}. 
It can be solved perturbatively via
a $d=2-\epsilon$ renormalization group expansion. Furthermore, 
all of the scaling exponents
can be obtained exactly in all $d$, owing to the existence of a 
``fluctuation-dissipation 
theorem" which we shall not elaborate here. One obtains the directed polymer
results $\zeta = 2/3$ and $\theta = 1/3$ which
 we have referred to several times in the
text. Not much is known analytically of the noisy-Burgers' equation
 away from 1+1 dimensions. Recently,
an uncontrolled mode-coupling approach~\cite{mc1,mc2} 
along the line of Ref.~\cite{HF} has been applied
to solve Eq.~(\ref{kpz}) in 2+1 dimensions, yielding exponent values 
that are rather close to
those obtained from numerical simulations~\cite{krug}. 
However, a controlled solution is
yet to be developed.

Although quantitative solution 
(e.g., the exponent values) of the noisy-Burgers' equation
are not available for $d \not= 1$, we can 
still learn much about the qualitative
behavior of the directed polymer, e.g., 
the rare fluctuations, by exploiting this mapping
to hydrodynamics, and especially by exploiting symmetries, as we now describe.
From the standard dynamic field theory formulation of Langevin dynamics 
upon averaging over $\eta$~\cite{msr,janssen}, one can
turn Eq.~(\ref{kpz}) into the free-energy functional
used in Eq.~(\ref{cZ}), 
\begin{equation}
\Xi[\widetilde{F},F] = e^{-S[\widetilde{F},F]},
\end{equation}
 where 
\begin{equation}
S[\widetilde{F},F] = \int dt d^d\vec{x} 
\left[ -D \widetilde{F}^2(\vec{x},t) + \widetilde{F}(\vec{x},t)
\left(\partial_t F(\vec{x},t) 
- \frac{T}{2\kappa}\vec{\nabla}^2 F + \frac{1}{2\kappa}
(\vec{\nabla} F)^2\right) \right].
\end{equation}
Since the disorder potential $\eta$ has already been averaged out,
translational symmetries in both $\vec{x}$ and $t$ are restored in 
the limit of large $t$, corresponding to the hydrodynamic steady state.
It is then convenient to go to  Fourier space, where
\begin{eqnarray}
&S[\widetilde{F},F] &= \int \frac{d\omega}{2\pi} \frac{d^d\vec{k} }{(2\pi)^d}
\left[ -D|\widetilde{F}(\vec{k},\omega)|^2 + 
\left(\frac{T}{2\kappa} k^2 -i\omega\right)
\widetilde{F}(-\vec{k},-\omega)F(\vec{k},\omega)\right] \label{S} \\
& &+ \int \frac{d\omega_1}{2\pi}
\frac{d\omega_2}{2\pi} \frac{d^d\vec{k}_1 }{(2\pi)^d}
\frac{d^d\vec{k}_2}{(2\pi)^d}
\left(-\frac{1}{2\kappa}\vec{k}_1\cdot\vec{k}_2\right)
\widetilde{F}(-\vec{k}_1-\vec{k}_2,-\omega_1-\omega_2)F(\vec{k}_1,\omega_1)
F(\vec{k}_2,\omega_2). \nonumber
\end{eqnarray}
The functional $S[\widetilde{F},F]$ is called the dynamic functional 
in the study of stochastic dynamics~\cite{janssen}.
In the absence of randomness, i.e., for $D=0$, this is just 
 the ``bare" generating functional of the vertex functions, 
\begin{equation} S[\widetilde{F},F] = \Gamma^{(0)}[\widetilde{F},F].
\end{equation}
From the definition of the vertex functions (\ref{vertex}), we have
\begin{equation}
 \Gamma[\widetilde{F},F] = \sum_{m,n} \frac{1}{m! n!} 
\int_{(\vec{k},\omega)}\widehat{\widehat{\Gamma}}_{m,n}\widetilde{F}^m F^n,
\end{equation}
where $\widehat{\widehat{\Gamma}}_{m,n}$ is the  Fourier transform in 
$\vec{x}$, $\vec{y}$
 and $t$ of the function $\gamma_{m,n}$ in Eq.~(\ref{vertex}). 
Comparing the above to Eq.~(\ref{S}), 
we can easily read off the bare vertex functions,
\begin{eqnarray}
& &\widehat{\widehat{\Gamma}}^{(0)}_{1,1}(\vec{k},\omega) = 
\frac{T}{2\kappa} k^2 - i \omega,
\label{gamma11} \\
& &\widehat{\widehat{\Gamma}}^{(0)}_{1,2}
(\vec{k}_1,\omega_1;\vec{k}_2,\omega_2)
 = -\frac{1}{\kappa}
\vec{k}_1 \cdot \vec{k}_2,\label{gamma12}\\
& &\widehat{\widehat{\Gamma}}^{(0)}_{2,0}
(\vec{k},\omega) = 0. \label{gamma20} 
\end{eqnarray}
The relevance of the random potential in the
 pinned phase makes the renormalized
functional $\Gamma[\overline{\widetilde{F}},
\overline{F}]$ much more complicated. However,
we will show in Appendix C that, as a consequence of the
statistical tilt symmetry, the {\it scaling} behavior obtained
from $\Gamma_{1,2}^{(0)}$ and $\Gamma_{1,2}$ are the same.
This permits us to use the bare vertex $\Gamma_{1,2}^{(0)}$ in place 
of the full vertex
$\Gamma_{1,2}$ to compute the tail of the 
 droplet distribution $W(\vec{\Delta},\tau)$ in Sec.III.A. 

As shown in Sec.III.A, $W(\vec{\Delta},\tau)$ is obtained by adding 
a source term
$\widetilde{J}$ to the random potential $\eta$, 
and then following the changes in $\overline{F}$.
In the context of stochastic dynamics, 
this procedure amounts to  probing the ``response" 
of the hydrodynamic field $\overline{F}$
to a small ``perturbation" $\widetilde{J}$ added to the right-hand side of 
Eq.~(\ref{kpz}).
We see that the distribution functions 
$\delta \overline{F}/\delta\widetilde{J} = G$ 
and $\delta^2 \overline{F}/\delta\widetilde{J}^2 = G_{2,1}$ are simply
the linear and nonlinear response functions of the noisy-Burgers' equation.
Furthermore, the {\it tails} of the distribution 
functions are similarly related to the 
hydrodynamic limits of the response functions. 
Thus the mapping to hydrodynamics
provides us with a convenient way of approaching the tails of distributions.

\newpage
\section{Statistical Tilt Symmetry and Ward Identities}

In this Appendix, we review the statistical tilt symmetry which
was used in Ref.~\cite{Schulz} to prove the 
result $\overline{\chi} = \tau/\kappa$.
The analogous symmetry in the noisy-Burgers' equation is the Galilean
invariance. It has been used in Refs.~\cite{FNS,KPZ} to derive the
exponent identity $\theta = 2\zeta - 1$. Here we shall derive the consequences
 of this symmetry, i.e., Ward identities,  more formally. The result is used 
to compute the droplet distribution in Sec.III.A.

We start with the Hamiltonian ${\cal H}[\vec{\xi},\eta]$ given in 
Eq.~(\ref{H}).
Consider the effect of an arbitrary bias of the form
\begin{equation}
{\cal H}_h[\vec{\xi},\eta] = 
{\cal H}[\vec{\xi},\eta] - \int_0^t dz \ \vec{h}(z)\cdot
\frac{d\vec{\xi}}{dz}. \label{H3}
\end{equation}
 By a simple coordinate transformation
$\vec{\xi}'(z) = \vec{\xi}(z) - \vec{r}(z)/\kappa$, 
where $\vec{r}(z) = \int^z dz'
\ \vec{h}(z')$, we obtain
\begin{equation}
{\cal H}_h[\vec{\xi},\eta] = {\cal H}[\vec{\xi}',\widetilde{\eta}]
- \frac{1}{2\kappa} \int_0^t dz' \ \vec{h}^2(z'),
\end{equation}
with a transformed random potential
\begin{equation}
 \widetilde{\eta}(\vec{\xi}',z) 
\equiv \eta[\vec{\xi}'(z)+ \vec{r}(z)/\kappa,\ z]. \label{C0}
\end{equation}
Similarly the free energy of the polymer with a point fixed at
$\vec{\xi}(t)=\vec{x}$ is given by 
\eqn{C1}{F_h(\vec{x},t;\eta) 
= F(\vec{x}-\vec{r}(t)/\kappa,t;\widetilde{\eta}) 
-\frac{1}{2\kappa} \int_0^t dz \ \vec{h}^2(z).}
Note that since the random potential $\eta$ is gaussian and uncorrelated,
the distribution of 
$\widetilde{\eta}$ is also gaussian, characterized by its covariance
\begin{eqnarray}
&\overline{\widetilde{\eta}(\vec{x}_1,t_1)\widetilde{\eta}(\vec{x}_2,t_2)}
&=\overline{\eta(\vec{x}_1,t_1)\eta(\vec{x}_2,t_2)} \nonumber \\
& &= 2D\delta^d(\vec{x}_1-\vec{x}_2)\delta(t_1-t_2) \label{C2}.
\end{eqnarray}
Thus disorder averages of a functional of $F$ 
over $\eta$ and $\widetilde{\eta}$ are identical. In particular,
if we choose $\vec{h}(z) = \vec{h}$ for $t_0 < z < t$,
and $\vec{h}(z) = 0$ for $ 0 < z < t_0$, then Eq.~(\ref{H3}) reduces to
Eq.~(\ref{H2}), and  
Eqs.~(\ref{C1}) and (\ref{C2}) imply that
the disorder-averaged linear response is just 
\begin{equation}
\overline{\chi} = - \frac{\partial}{\partial h_i^2} 
\overline{F}_h(\vec{x},t)\Bigg|_{\vec{h}=0} =\frac{t-t_0}{\kappa}.
\end{equation}
This result was first proved for a class of disordered systems 
in Ref.~\cite{Schulz}.

In the following, we shall exploit this statistical symmetry  more formally
by considering in detail the case $t_0=0$, corresponding to applying a
 uniform tilt field $\vec{h}$. In this case, Eqs.~(\ref{C0}) and (\ref{C1})
become
\begin{equation}
 \widetilde{\eta}(\vec{\xi}',z) 
\equiv \eta[\vec{\xi}'(z)+ \vec{h}z/\kappa,\ z], \label{C0'}
\end{equation}
and
\begin{equation}
F_h(\vec{x},t;\eta) = F(\vec{x}-\vec{h}t/\kappa,\ t;\widetilde{\eta}) 
-\frac{\vec{h}^2}{2\kappa} t.\label{C1'}
\end{equation}
 From Eq.~(\ref{C1'}), it is clear that $F_h$ no longer satisfies the
noisy-Burgers' equation (\ref{kpz}). However, the combination
$F_h(\vec{x},t) - T\vec{h}\cdot \vec{x}$ does. This is because
 Eq.~(\ref{kpz}) is invariant under the ``Galilean'' transformation
\begin{equation}
F'(\vec{x},t) = F(\vec{x} + \vec{v} t, t) - \kappa
\vec{v} \cdot \vec{x},
\end{equation}
and 
\begin{equation}
\eta'(\vec{x},t) = \eta(\vec{x}+\vec{v} t,t),
\end{equation}
as can be verified straightforwardly for $\vec{v} = \vec{h}/\kappa$.
We now use this Galilean invariance to derive some useful Ward identities.

The statistical tilt symmetry implies that the generating functional of the 
vertex function $\Gamma[\overline{\widetilde{F}},\overline{F}]$
is invariant under the Galilean transformation of the disorder-averaged fields
$\overline{F}$ and $\overline{\widetilde{F}}$. 
In Fourier space, the transformation reads,
\begin{eqnarray}
&\overline{F}'(\vec{q},t_1) &= e^{i\vec{v}\cdot\vec{q} t_1}
 \overline{F}(\vec{q},t) + i
\kappa \vec{v}\cdot\vec{\nabla}_{\vec{q}} 
\delta^d(\vec{q}), \label{GI1} \\
&\overline{\widetilde{F}}'(\vec{q},t_1) 
&= e^{i\vec{v}\cdot\vec{q} t} 
\overline{\widetilde{F}}(\vec{q},t_1). \label{GI2} 
\end{eqnarray}
Thus we have the identity
\eqn{GI0}{\frac{\delta\Gamma}{\delta\vec{v}} = 0 = \int d^d\vec{q}dt_1 
\left\{ i \left[\vec{q} t_1 \overline{F}(\vec{q},t_1)- \kappa 
\delta^d(\vec{q})\vec{\nabla}_{\vec{q}}\right] 
\frac{\delta\Gamma}{\delta\overline{F}(\vec{q},t_1)}
+ i \vec{q} t_1 \overline{\widetilde{F}}
(\vec{q},t_1)\frac{\delta\Gamma}{\delta\overline{\widetilde{F}}(\vec{q},t_1)}
\right\},}
where Eqs.~(\ref{GI1}) and (\ref{GI2}) have been 
used to obtain the variations of $\overline{F}$ 
and $\overline{\widetilde{F}}$
with respect to $\vec{v}$. Taking functional 
derivatives of Eq.~(\ref{GI0}) with
respect to $\overline{\widetilde{F}}(\vec{k},t)$ 
and $\overline{F}(\vec{k}-\vec{q},t_2)$, then taking the limit
$\overline{\widetilde{F}}, \overline{F} \to 0$, 
we obtain the following Ward identity in Fourier space,
\eqn{Ward}
{ \kappa \vec{\nabla}_{\vec{q}}\int dt_1\widehat{\Gamma}_{1,2}(\vec{q},t-t_1;
\vec{k}-\vec{q},t-t_2)|_{\vec{q}\to 0} 
= \vec{k} (t-t_2) \widehat{\Gamma}_{1,1}(\vec{k},t-t_2),}
where the definition
\begin{eqnarray}
&\widehat{\Gamma}_{1,n}&(\vec{q}_1,t-t_1;\cdots;\vec{q}_n,t-t_n)
(2\pi)^{d}\delta^d(\vec{q}_1+\cdots+\vec{q}_n+\vec{k}) \nonumber \\
& &\equiv \int d^d\vec{x} d^d\vec{x}_1\cdots d^d\vec{x}_n 
e^{i\vec{k}\vec{x} + i\vec{q}_1\vec{x}_1 + \cdots + i\vec{q}_n\vec{x}_n}
\frac{\delta^{(1+n)}\Gamma[\overline{\widetilde{F}},
\overline{F}]}{\delta\overline{\widetilde{F}}(\vec{x},t)
\delta\overline{F}(\vec{x}_1,t_1) 
\cdots \delta\overline{F}(\vec{x}_n,t_n)} \nonumber
\end{eqnarray}
is used. It is more convenient to write Eq.~(\ref{Ward}) in frequency space.
With 
\begin{eqnarray}
&\widehat{\widehat{\Gamma}}_{1,n}
&(\vec{q}_1,\omega_1;\cdots;\vec{q}_n,\omega_n)  
2\pi \delta(\omega_1 + \cdots + \omega_n + \omega)\nonumber\\
& & \equiv 
\int dt dt_1 \cdots dt_n e^{-i\omega t - i \mu_1 t_1  \cdots - i\mu_n t_n}
\widehat{\Gamma}_{1,n}(\vec{q}_1,t-t_1;\cdots;\vec{q}_n,t-t_n),\nonumber 
\end{eqnarray}
Eq.~(\ref{Ward}) becomes
\eqn{Ward2}{\kappa \vec{\nabla}_{\vec{q}}
\widehat{\widehat{\Gamma}}_{1,2}(\vec{q},\mu;
\vec{k}-\vec{q},\omega-\mu)|_{\vec{q},\mu\to 0} 
= - i \vec{k} \frac{\partial}{\partial\omega}
\widehat{\widehat{\Gamma}}_{1,1}(\vec{k},\omega).}

From Eq.~(\ref{cG11}), we have 
\begin{equation}
\widehat{\widehat{\Gamma}}_{1,1}(\vec{k},\omega)
 = \widehat{\widehat{G}}^{-1}(\vec{k},\omega), \label{propagator}
\end{equation}
where 
\begin{equation}
\widehat{\widehat{G}}(\vec{k},\omega)
 = \int dt \widehat{g}(qt^\zeta) e^{-i\omega t}.
\end{equation}
In the simplest case where there is no disorder, we  have
\begin{equation}
\widehat{\widehat{\Gamma}}_{1,1}^{(0)}(\vec{k},\omega) 
= \frac{T}{2\kappa} k^2 - i \omega.
\end{equation}
The Ward identity then reads 
\begin{equation}
\kappa \vec{\nabla}_{\vec{q}}
\widehat{\widehat{\Gamma}}_{1,2}^{(0)}(\vec{q},\mu;
\vec{k}-\vec{q},\omega-\mu)|_{\vec{q},\mu\to 0} = - \vec{k},
\end{equation}
yielding a ``bare" three-point vertex 
\eqn{bare2}{\widehat{\widehat{\Gamma}}_{1,2}^{(0)}(\vec{k}_1,\omega_1;
\vec{k}_2,\omega_2) = -\frac{1}{\kappa} \vec{k}_1\cdot\vec{k_2},} since
$\widehat{\widehat{\Gamma}}_{1,2}$ must 
be symmetric in $\vec{k}_1$ and $\vec{k_2}$. 
This vertex is exactly Eq.~(\ref{gamma12}), which has been obtained directly 
from the dynamic functional $S[\widetilde{F},F]$ in Appendix B.

For the random problem at hand,
 the scaling form Eq. (\ref{Pinfinite}) or (\ref{hG}) for $G$  
implies that the two-point vertex must have the following form:
\begin{equation}
\widehat{\widehat{\Gamma}}_{1,1}(\vec{k},\omega) 
= k^{1/\zeta}\widetilde{\Gamma}_{1,1}(\omega/k^{1/\zeta}),
\end{equation}
where
\begin{equation}
\widetilde{\Gamma}_{1,1}(s) = \nu_1(s) - i s \nu_2(s), \label{tg11}
\end{equation}
with $\nu_1$, $\nu_2$ being real, {\it dimensionless} scaling functions.
Since $\widehat{g}$ is real, $\nu_1$ and $\nu_2$ must be even functions.
Furthermore, $\int_0^\infty dt \widehat{g}(qt^\zeta) \sim q^{-1/\zeta}$ since
$\widehat{g}$ is rapidly decreasing for large argument. It then follows that
\begin{equation}
\lim_{s\to 0} \nu_1(s) = {\rm const} \qquad {\rm and} \qquad
\lim_{s\to 0} s \ \nu_2(s) = 0. \label{cond}
\end{equation}
The Ward identity (\ref{Ward2}) now leads to the relation
\begin{equation}
\widehat{\widehat{\Gamma}}_{1,2}
(\vec{q},\mu;\vec{k}-\vec{q},\omega-\mu)|_{\vec{q},\mu\to 0} 
 = -\frac{1}{\kappa} \vec{q}\cdot(\vec{k}-\vec{q}) \nu(\omega/k^{1/\zeta}),
\end{equation}
 where $\nu(s) =  i
\frac{\partial}{\partial s}\widetilde{\Gamma}_{1,1}(s)$ is another
 dimensionless (but complex) scaling function,
\begin{equation}
\nu(s) = \nu_2(s) + s \frac{d\nu_2}{ds} 
		- i \frac{d\nu_1}{ds}. \label{nu}
\end{equation}
The above limiting behavior of $\widehat{\widehat{\Gamma}}_{1,2}$
 is as much as the Ward identity
will directly yield. However, 
it does {\it suggest} a natural scaling form for the
full spatial and temporal dependence of the vertex,
\eqn{hhG12}{\widehat{\widehat{\Gamma}}_{1,2}
(\vec{k}_1,\omega_1;\vec{k}_2,\omega_2)
 = -\frac{1}{\kappa} \vec{k}_1 \cdot \vec{k}_2 
V(\omega_2/k_2^{1/\zeta},k_1/k_2,\omega_1/\omega_2),}
where $V$ is another complex, {\it dimensionless}
 scaling function, with $V(s,0,0) = \nu(s)$.

The scaling form Eq.~(\ref{hhG12}) of the 
full vertex $\widehat{\widehat{\Gamma}}_{1,2}$,
together with mild convergence conditions on $V$, imply that the {\it scaling}
properties obtained from the full vertex is the same as that obtained from 
the bare vertex (\ref{bare2}).
We now illustrate this by computing the tail of the droplet distribution,
or equivalently the singularity in 
$\widehat{W}'(\vec{q},\tau\to\infty)$, using the
full vertex.

In the limit $\tau\to \infty$, it is convenient to write Eq.~(\ref{hW}) in
frequency space,
\begin{eqnarray}
&\widehat{W}'(\vec{q},t\to\infty) 
&= -T \int_{-\infty}^\infty \frac{d\omega}{2\pi} 
\ \widehat{\widehat{\Gamma}}_{1,2}
(\vec{q},\omega;-\vec{q},-\omega) \widehat{\widehat{G}}(\vec{q},\omega)
\widehat{\widehat{G}}(-\vec{q},-\omega)\nonumber \\
& &= -\frac{T}{\kappa} q^{2-1/\zeta} I,
\end{eqnarray}
with
\begin{equation}
I = \int_{-\infty}^\infty \frac{ds}{2\pi}
\frac{V(s,1,1)}{\nu_1^2(s) + s^2 \nu_2^2(s)},
\end{equation}
where we used Eqs.~(\ref{tg11}) and (\ref{hhG12}).
The scaling form Eq.~(\ref{cusp}) is valid as long as the integral $I$
converges. In Sec.III.A, we used the ``bare'' $\Gamma$'s corresponding to
$V=\nu_1=\nu_2 = 1$. This gives the result $I=1/2$. In general, we see
that, given the properties of $\nu_1$ and $\nu_2$ in Eq.~(\ref{cond}),
 the integral converges if
\begin{equation}
V(s\to 0,1,1) < s^{-1} \qquad {\rm and} \qquad 
V(s\to \infty,1,1) < s \ \nu_2^2(s). \label{conver}
\end{equation}
Unfortunately, $V(s,1,1)$ is not directly obtainable from the Ward identity,
which only gives $V(s,0,0)$. However, we do not expect $V(s, k_1/k_2,
\omega_1/\omega_2)$ to depend strongly  on the ratio $k_1/k_2$ and
$\omega_1/\omega_2$ in the region $0\le k_1/k_2 \le 1$ and $0 \le
\omega_1/\omega_2 \le 1$. If we use $V(s,0,0) = \nu(s)$ for $V(s,1,1)$ with
$\nu(s)$ given by Eq. (\ref{nu}),
then  the convergence conditions (\ref{conver}) are always satisfied
given the properties of $\nu_1$ and $\nu_2$ in Eq.~(\ref{cond}).
 It is thus very reasonable to 
expect that (\ref{conver}) are also satisfied by $V(s,1,1)$ itself,
which then leads to the scaling result Eq.~(\ref{cusp})
 for $\widehat{W}'(\vec{q},\tau)$, with $\widehat{w}(s\to\infty) = I$.

Finally we note that although the scaling form (\ref{hhG12}) for the vertex
and the convergence conditions (\ref{conver}) have not been derived
from first principles, they are very plausible if
an underlying renormalization group exists.
This is especially true in 1+1 dimensions, where the strong
coupling fixed point of the noisy-Burgers' equation
can be described perturbatively  
via a $d=2-\epsilon$ renormalization-group expansion of the anisotropic
 driven diffusion equation (\ref{dds}) 
in $d+1$ dimensions as explained in Appendix B.
The vertex function $\widehat{\widehat{\Gamma}}_{1,2}$ 
can be computed explicitly in 
the $2-\epsilon$ expansion, and is indeed described by the scaling
form Eq.~(\ref{hhG12}). Away from 1+1 dimensions 
for the noisy-Burgers' equation, 
a perturbative renormalization scheme is 
not known. However none of the known results give any indication of 
the break down of simple scaling. Thus we expect 
that the hypothesized scaling form
(\ref{hhG12}) for $\widehat{\widehat{\Gamma}}_{1,2}$
 is valid in all dimensions
for the Burgers' equation and therefore also for the directed polymer.

\begin{figure}
\vspace*{2.0truein}
\epsfxsize=6.5truein 
\epsffile{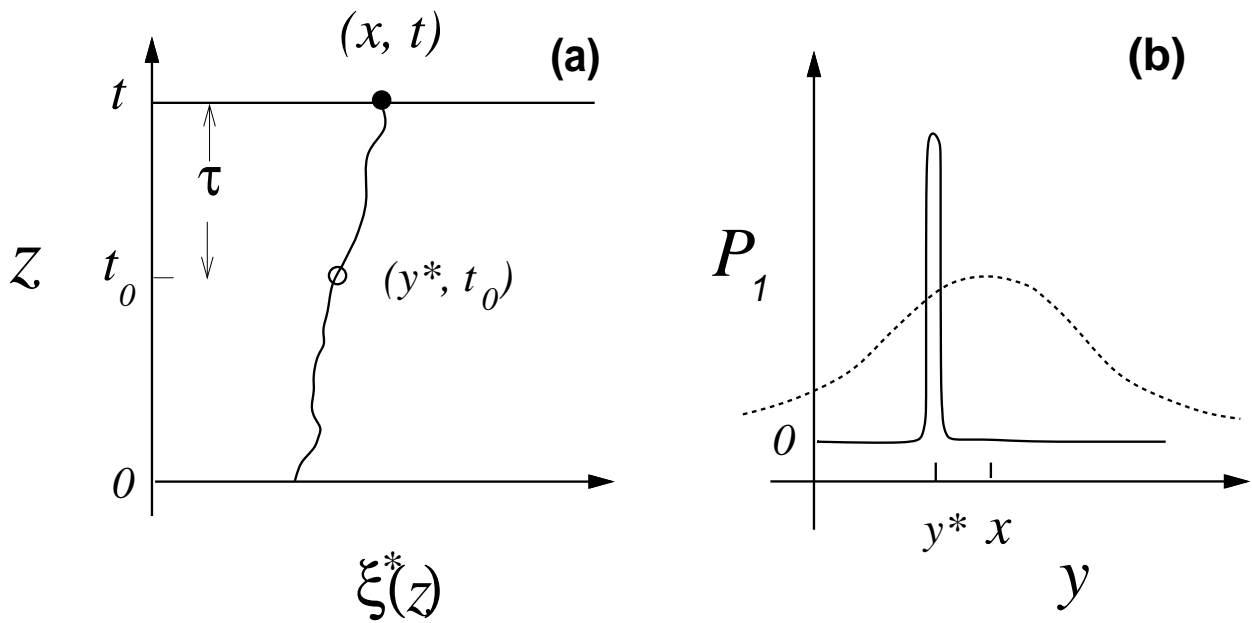}
\vskip 1in
\caption{ (a) Schematic of the ground state $\vec{\xi}^*(z)$ of a typical 
directed polymer with one end fixed at
$(\vec{x},t)$ (the solid circle). The position of a segment at $z=t_0$
 (open circle) is at $ \vec{\xi}^*(t_0) = \vec{y}^*$.  (b)~The one-point 
distribution function $P_1(\vec{y},t_0|\vec{x},t)$ as a function of $\vec{y}$.
It is sharply peaked near some sample-specific $\vec{y}^*$ 
with $|\vec{y}^*-\vec{x}|
\sim \tau^\zeta$, where $\tau = t- t_0$. The dashed
line indicates the disorder-averaged distribution function
$G_t(\vec{x}-\vec{y},\tau) = \overline{P}_1(\vec{y},t_0|\vec{x},t)$, which
is centered about $\vec{x}$, with a width $\tau^\zeta$.
This gives the distribution of $\vec{y}^*$ from sample-to-sample. }
\end{figure}

\begin{figure}
\vspace*{2.0truein}
\epsfxsize=6.5truein 
\epsffile{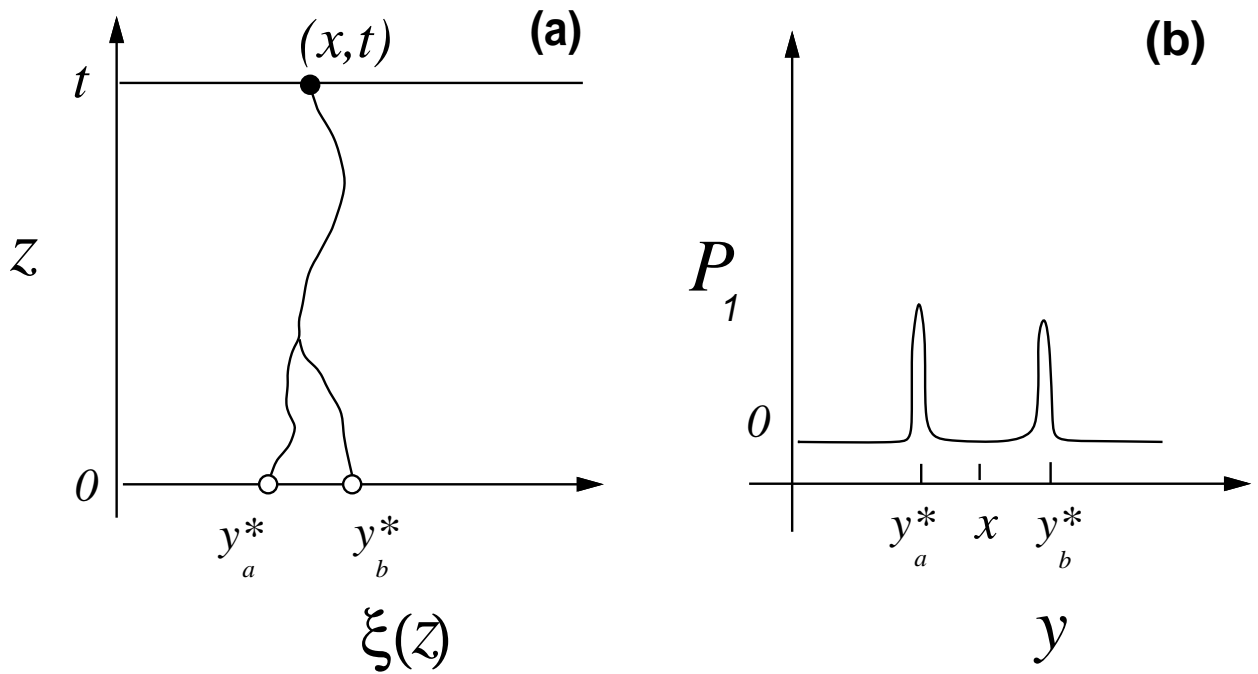}
\vskip 1in
\caption{ (a) Behavior with a configuration of 
the random potential that gives rise to
two almost-degenerate ``ground
states'' $\vec{\xi}_a^*(z)$ and $\vec{\xi}_b^*(z)$ for a directed polymer
 with one end
fixed at $(\vec{x},t)$. The positions of the free ends of the two states are
$\vec{\xi}_a^*(0)=\vec{y}_a^*$ and $\vec{\xi}_b^*(0)=\vec{y}_b^*$.
(b) The end-point distribution function 
$P_1(\vec{y},0|\vec{x},t)$ for the sample in 
(a) displays two peaks at $\vec{y}_a^*$
and $\vec{y}_b^*$.}
\end{figure}

\begin{figure}
\vspace*{2.0truein}
\hspace*{0.6truein}
\epsfxsize=4truein 
\epsffile{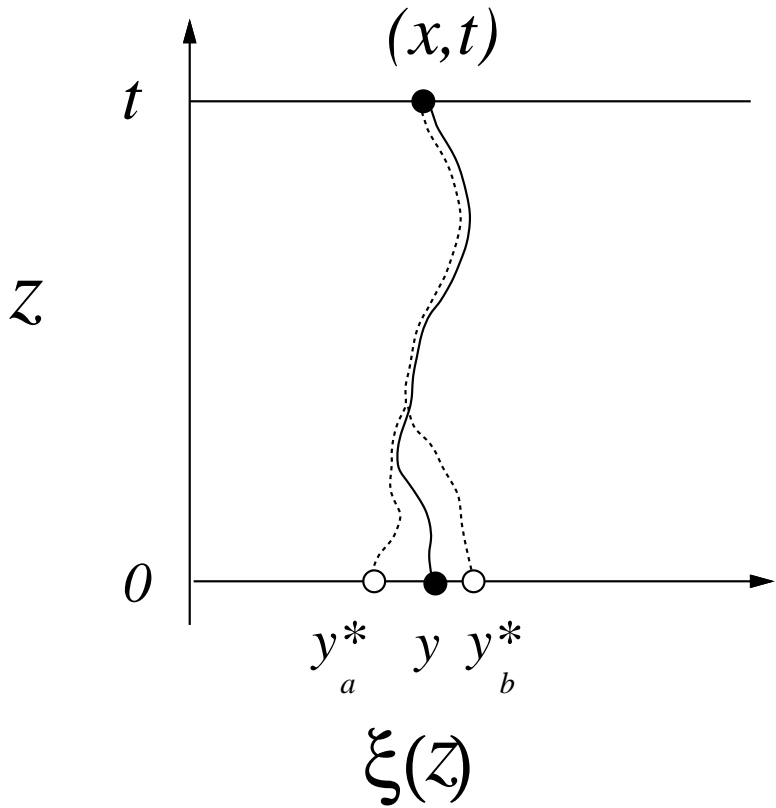}
\vskip 1in
\caption{ The dotted lines are the degenerate ground states 
$\vec{\xi}_a^*(z)$ and $\vec{\xi}_b^*(z)$ shown in Fig.~2(a). The solid line
is the optimal path with {\it both} ends fixed:
$\vec{\xi}(t)=\vec{x}$ and $\vec{\xi}(0)=\vec{y}$, where $\vec{y}$ is
between $\vec{y}_a^*$ and $\vec{y}_b^*$, 
which are separated by a distance of order
$\Delta$.}
\end{figure}

\begin{figure}
\vspace*{2.0truein}
\hspace*{0.6truein}
\epsfxsize=4truein 
\epsffile{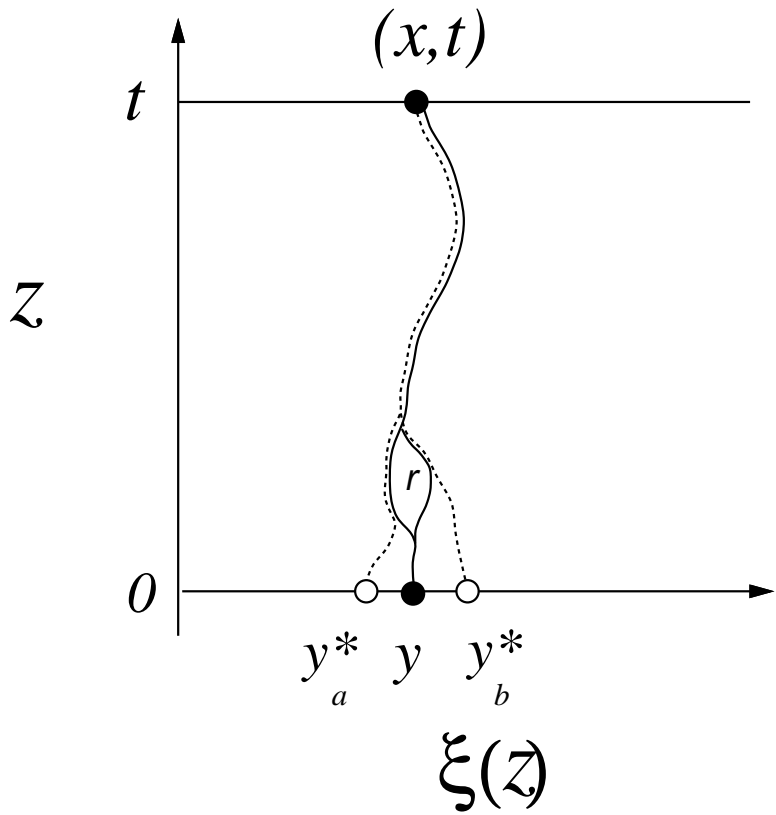}
\vskip 1in
\caption{ The solid line illustrates a possible degenerate configuration 
of size $r$ in the bulk for the optimal 
path shown in Fig.~3. Some energy barrier
separates the two degenerate pieces of the optimal paths to $\vec{y}$.
These will affect the energy barrier to move the polymer from between 
its almost-degenerate ground states ending at $y_a^*$ and $y_b^*$. }
\end{figure}

\begin{figure}
\vspace*{2.0truein}
\hspace*{0.6truein}
\epsfxsize=4truein 
\epsffile{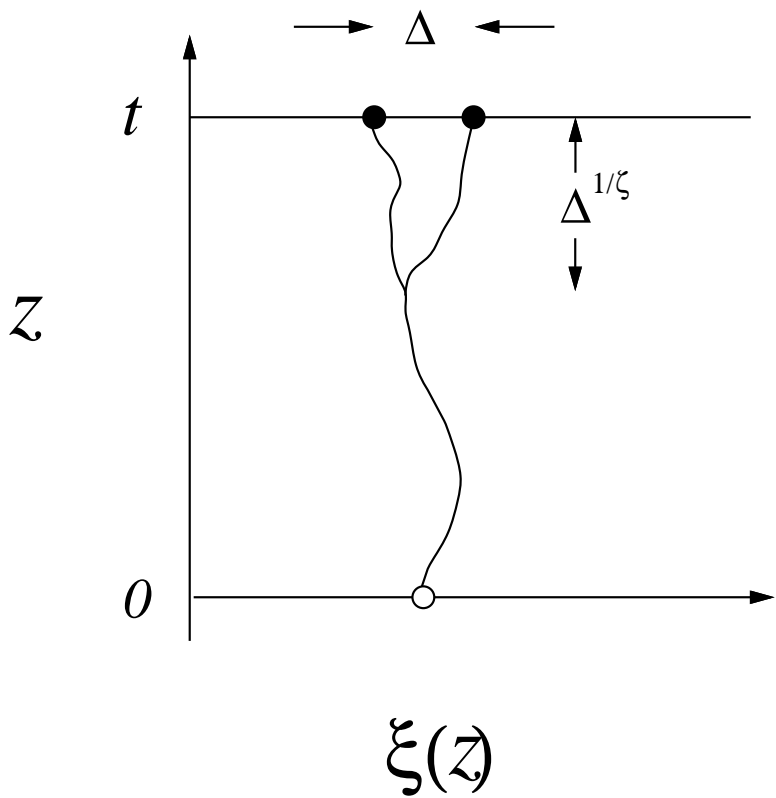}
\vskip 1in
\caption{Two polymers with ends fixed at 
a distance $\Delta$ apart typically merge
at a distance $\Delta^{1/\zeta}$ from the fixed ends. }
\end{figure}

\begin{figure}
\vspace*{2.0truein}
\hspace*{1.0truein}
\epsfxsize=4truein 
\epsffile{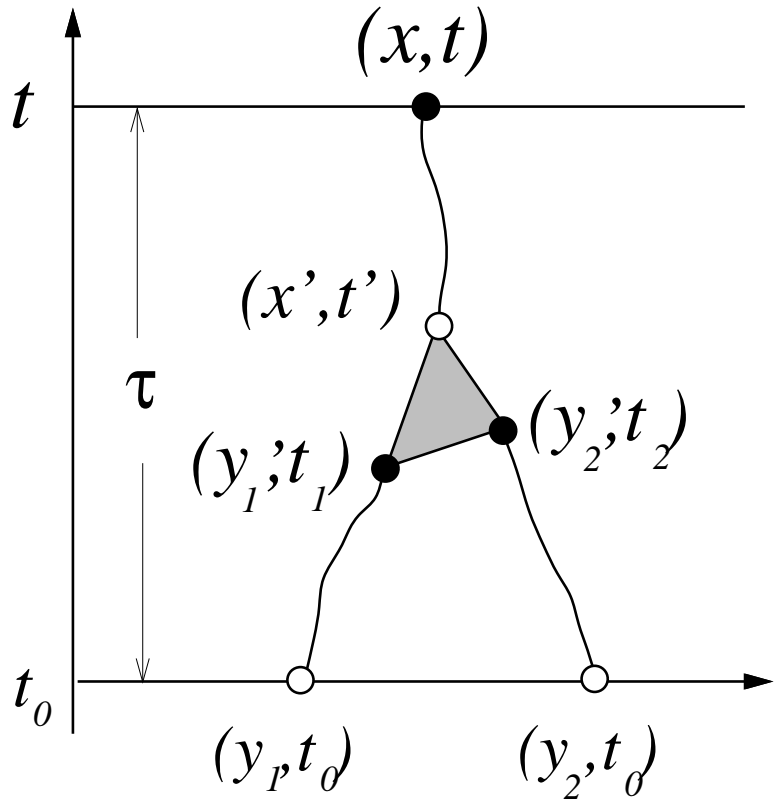}
\vskip 1in
\caption{ The three-point function $G_{2,1}$ can be obtained as the
convolution of three one-point functions $G$'s and a vertex function
$\Gamma_{1,2}$ (the shaded triangle)
which controls the branching process.}
\end{figure}

\begin{figure}
\vspace*{2.0truein}
\epsfxsize=5.5truein 
\epsffile{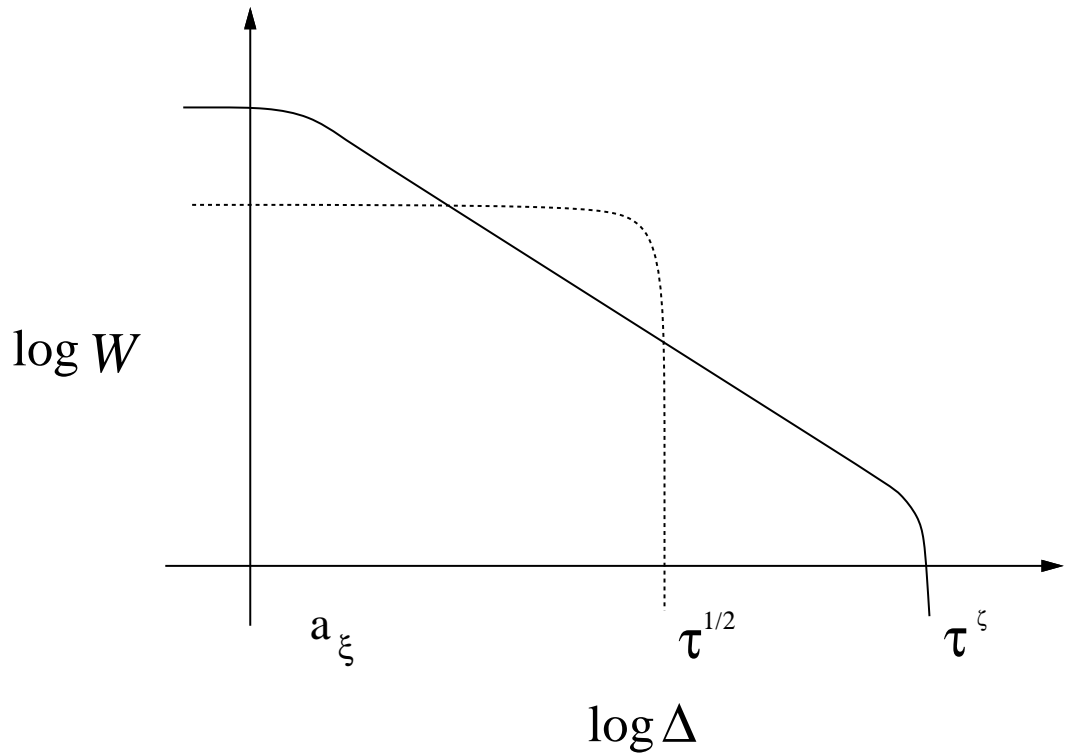}
\vskip 1in
\caption{ The distribution function of the active droplet of size $\Delta$,
$W(\vec{\Delta},\tau)$ (the solid line), is composed of two parts: A smooth
piece for $\Delta \le a_\xi(T)$, the small scale cutoff, 
and a long power-law tail
extending to $\tau^\zeta$.  The dashed line is the gaussian
distribution $W^{(0)}(\vec{\Delta},\tau)$ which describes the pure system 
(and the high temperature phase of the disordered system). 
It has a width of $O(\tau^{1/2})$. }
\end{figure}

\begin{figure}
\vspace*{2.0truein}
\hspace*{1.0truein}
\epsfysize=4.5truein 
\epsffile{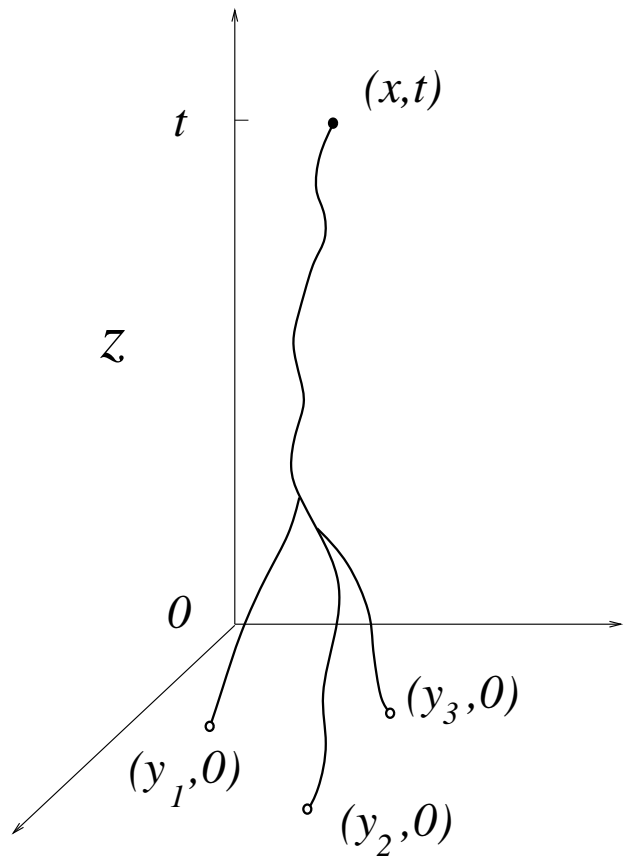}
\vskip 1in
\caption{ A triply degenerate ground state configuration for a directed
polymer in 2+1 dimensions, with the distance between each pair of end points
being of order $t^\zeta$ for a polymer of length $t$. }
\end{figure}


\begin{references}
\bibitem{twocopies} Alternatively, we can place two noninteracting polymers, 
$\vec{\xi}_a(z)$ and  $\vec{\xi}_b(z)$, in the {\it same}
random potential, both with the end $\vec{\xi}(t)$ fixed at $\vec{x}$.
The disorder-averaged  distribution 
$W_t(\vec{\Delta},\tau)$ is just the average probability
that the two polymers are separated 
by a distance $\vec{\Delta}$ at $z=t-\tau$,
which only occurs when the sample has two nearly degenerate ground states
that are located away from each other by  
a displacement $\vec{\Delta}$ at $z$.
\bibitem{offdiagonal} The behavior of the off-diagonal response,
$$ \chi_{ij}[\eta] = \frac{\partial}{\partial h_j} 
\avg{\xi_i(0)}_{\vec{h}\to 0}, $$
with $i \not= j$ is also interesting and nontrivial.
But we will not pursue this here
in order to focus the study on the simplest rare fluctuations.
\bibitem{note2} Rare events are also uncovered in the
replica-symmetric solution of the directed polymer
 in $1+1$ dimensions~\cite{p}, where a free-energy
functional of the type described in Appendices A and B is used.
\bibitem{cavity} Our analysis based
on the ``evolution" of the free energy for 
successively larger systems somewhat
resembles the ``cavity method'' introduced 
in the study of mean-field spin glasses
\cite{sg}.
\end{references}
\end{document}